\documentclass[english]{article}
\usepackage{times}
\usepackage{url}
\usepackage{bbm}
\usepackage[T1]{fontenc}
\usepackage[latin1]{inputenc}
\usepackage{geometry}
\usepackage{rotating}
\usepackage{color}
\usepackage{graphicx}
\usepackage{amsmath, amsthm, amssymb}
\usepackage{setspace}
\usepackage{lineno}
\usepackage{hyperref}
\usepackage{natbib} \bibpunct{(}{)}{;}{author-year}{}{,} 

\usepackage{color}
\definecolor{trustcolor}{rgb}{0,0,1}

\makeatletter

\usepackage{babel}
\makeatother
\begin{document}

\title{Genome scans for detecting footprints of local adaptation using a Bayesian factor model}

\author{Nicolas Duforet-Frebourg$^{1}$, Eric Bazin$^{2}$, Michael G.B. Blum$^{1}$}

\date{~ }

\maketitle
\noindent$^{1}$ Universit\'e Joseph Fourier, Centre National de la Recherche Scientifique, Laboratoire TIMC-IMAG, UMR 5525, Grenoble, France.\\
$^{2}$ Universit\'e Joseph Fourier, Centre National de la Recherche Scientifique, Laboratoire d'Ecologie Alpine, UMR 5553, Grenoble, France.


\noindent Running Head:  Genome scans based on Bayesian factor model 

Corresponding author:
Michael  Blum

Laboratoire TIMC-IMAG, Facult\'e de M\'edecine, 38706 La Tronche, France 

Phone +33 4 56 52 00 65

Fax +33 4 56 52 00 55

Email: michael.blum@imag.fr

\clearpage
\begin{abstract}
{\normalsize 
A central part of population genomics consists of finding genomic regions implicated in local adaptation. Population genomic analyses are based on genotyping numerous molecular markers and looking for outlier loci in terms of patterns of genetic differentiation. One of the most common approach for selection scan is based on statistics that measure population differentiation such as $F_{ST}$. However they are important caveats with approaches related to $F_{ST}$ because they require grouping individuals into populations and they additionally assume a particular model of population structure. Here we implement a more flexible individual-based approach based on Bayesian factor models. Factor models capture population structure with latent variables called factors, which can describe clustering of individuals into populations or isolation-by-distance patterns. Using hierarchical Bayesian modeling, we both infer population structure and identify outlier loci that are candidates for local adaptation. As outlier loci, the hierarchical factor model searches for loci that are atypically related to population structure as measured by the latent factors. In a model of population divergence, we show that the factor model can achieve a 2-fold or more reduction of false discovery rate compared to the software BayeScan or compared to a $F_{ST}$ approach.  We analyze the data of the Human Genome Diversity Panel to provide an example of how factor models can be used to detect local adaptation with a large number of SNPs. The Bayesian factor model is implemented in the open-source PCAdapt software.}
\end{abstract}

\clearpage

\section*{Introduction}
With the development of sequencing and genotyping technologies, there is a considerable impetus to pinpoint loci involved in local adaptation \cite[]{akey02,bonin06}.
Finding genomic regions subject to local adaptation is a central part of ``population genomics'', which is based on genotyping numerous molecular markers and looking for outlier loci \cite[]{luikart03}. The main principle of population genomics is that most loci have neutral patterns of variation that are similarly affected by demographic processes whereas loci targeted by natural selection have atypical patterns. Measures of genetic differentiation between populations such as $F_{ST}$  have been commonly used to find outlier loci although there are many alternative approaches \cite[]{oleksyk10}. Loci involved in local adaptation have increased values of genetic differentiation for populations living in different environments.  A proof of concept was provided when studying human adaptation to altitude because the most differentiated variants between a Tibetan population living in a hypoxic environment and a lowland Han Chinese population were found in hypoxia-inducible transcription factors \cite[]{yi2010,xu11}.

Genome scans based on $F_{ST}$ were proposed by \cite{lewontin73} and have been considerably expanded since \cite[]{beaumont96,vitalis01,beaumont04,foll08,riebler08,guo09,bonhomme10,bazin10,gompert11,fariello13}.
They are not limited to two populations as in the adaptation-to-altitude example and can be used with multiple populations. One possibility is to compute an overall $F_{ST}$ measure of genetic differentiation and to determine a threshold at which the null hypothesis of neutral evolution can be rejected \cite[]{beaumont96}. Another possibility is to adopt a Bayesian perspective by implementing the multinomial-Dirichlet model or F-model, which is parametrized by population-specific $F$-statistics \cite[]{beaumont04}. The $F$-statistics can be interpreted as measures of divergence from a common immigrant gene pool \cite[]{wright31} or as as divergence from an initial and  hypothetical ancestral population \cite[]{nicholson02}. The Bayesian approach for distinguishing between neutral or adaptive evolution offers the opportunity to assign a probability to each of the two evolutionary models at each locus \cite[]{foll08,riebler08}. In the following, we refer to genome scans based on $F$-statistics for approaches based on $F_{ST}$ or based on population-specific $F$-statistics. 
There are many software implementing genome scans based on $F$-statistics (e.g. {\it BayeScan, DetSel, fdist2, Lositan}), and they contribute to the popularity of this approach in population genomics \cite[]{beaumont96,vitalis03,antao08,foll08}.

However, a major issue with genome scans based on $F$-statistics is that they can generate a high rate of false-positives for both biological and statistical reasons \cite[]{bierne13}.  Here we propose to address the statistical and computational problems that arise with $F$-statistics. The first problem arises because $F$-statistics have been derived under the Wright's $F$-model of population subdivision, which assumes a particular covariance structure for gene frequencies among populations \cite[]{bierne13,fourcade13}. When spatial structure  departs from Wright's island model of population subdivision, genome scans based on $F$-statistics produce many false positives  and alternative statistical measures that account for population structure have recently been proposed \cite[]{bonhomme10,gunther13}. A second potential problem concerns the computational burden of some Bayesian approaches, which can become an obstacle with large number of SNPs \cite[]{lange2014}. The last intrinsic problem of genome scans based on $F$-statistics is that individuals should be grouped into populations. However, it has been advocated in landscape genetics to rather work at the scale of individuals because it avoids potential bias in identifying populations in advance and it offers the opportunity to conduct studies at a finer scale \cite[]{manel13}.

To tackle the aforementioned problems, we propose a statistical method based on a Bayesian factor model \cite[]{west03} to pick outlier loci involved in local adaptation. With factor models, we seek to jointly determine population structure and outlier loci. Factor models are strongly related to principal components analysis (PCA) because they both approximate the matrix of individual genotypes by a product of two lower-rank matrices, albeit using different constraints and priors for the lower-rank matrices \cite[]{engelhardt10}. One of the two matrices encodes population structure using latent factors whereas the second matrix measures to what extent each individual SNP is related to the pattern of population structure. The proposed factor model seeks for loci that are atypically related with population structure. To show the potential of factor models for genome scan, we consider two examples. First, we consider a model of population divergence. In this example, we compare  false discovery rates obtained with the proposed factor model, with {\it BayeScan}, and with a genome-scan based on $F_{ST}$. The second example is a model of isolation-by-distance with selection. It is an instance of how factor models can be used to detect local adaptation when it would be arbitrary to group individuals into populations. Last, we analyze the HGDP human data-set \cite[]{li08} to provide an example of how factor models can be used to detect local adaptation with a large number of SNPs.

\section*{New Approaches}
We denote by ${\bf G}$ the $n\times p$ matrix of allele counts where $n$ is the number of individuals and $p$ is the number of loci. The elements $G_{i\ell}$, $i=1,\dots,n$, $\ell=1,\dots,p$, correspond to the allele counts of the $i^{th}$ individual at locus $\ell$ and belong to $\{0,1\}$ or $ \{0,1,2 \}$ for haploid and diploid species respectively.  We assume that the matrix of genotypes has been centered (each column has a mean of 0), and the resulting matrix is denoted by ${\bf Y}$.

Factor models assume that the matrix of column-centered genotypes ${\bf Y}$ can be written as a product of two lower-rank matrices ${\bf U}$ and ${\bf V}$ of dimension $(n\times K)$ and $(K\times p)$ where $K$ is an hyper parameter of the factor model. Denoting by ${\bf U}_1$,\dots, ${\bf U}_K$ the column-vectors of ${\bf U}$ referred as {\it factors} or {\it latent factors} in the following, factor models assume that the vector--of size $n$--of centered allele counts ${\bf Y}_{\ell}$ can be obtained as 
\begin{equation}
\label{eq:factor}
{\mathbf Y}_{\ell}=\sum_{k=1}^K {\mathbf U}_{k} {V_{k\ell}} + {\boldsymbol \epsilon}_{\ell},\; \ell=1\,\dots,p,
\end{equation}
where ${\boldsymbol \epsilon}_{\ell}$  is a vector containing $n$ independent Gaussian residuals of variance $\sigma^2$ and where the $V_{k\ell}$ are the elements of the matrix ${\bf V}$. Assuming that the $K$ factors are known, then the elements $V_{k\ell}$ of the matrix ${\bf V}$ are the regression coefficients obtained after regressing  the vector of centered allele counts ${\bf Y}_{\ell}$ by the $K$ factors ${\bf U}_1$,\dots ${\bf U}_K$. The outlier approach we advocate is to consider as candidates for local adaptation the loci $\ell$ that have large (in absolute value) regression coefficients ${\bf V}_{k\ell}$ for one of the factor ${\bf U}_1$,\dots, ${\bf U}_K$. In factor models, the $K$ factors ${\bf U}_1$,\dots, ${\bf U}_K$ are in fact unknown and have to be estimated; they are parameters of the model and represent population structure \cite[]{engelhardt10}. In our proposed framework, outlier loci are therefore loci that are excessively related with population structure, as measured by the latent factors. After statistical inference, the factors ${\bf U}_1$,\dots, ${\bf U}_K$ are ordered by decreasing variances  $\sigma_1^2>\dots>\sigma_K^2$ where  $\sigma_k^2$ measures the variance of the regression coefficients ${\bf V}_{k\ell}$ for the $k^{\rm th}$ factor.

\begin{figure}[ht!]
	\centering
		\includegraphics[width=14cm]{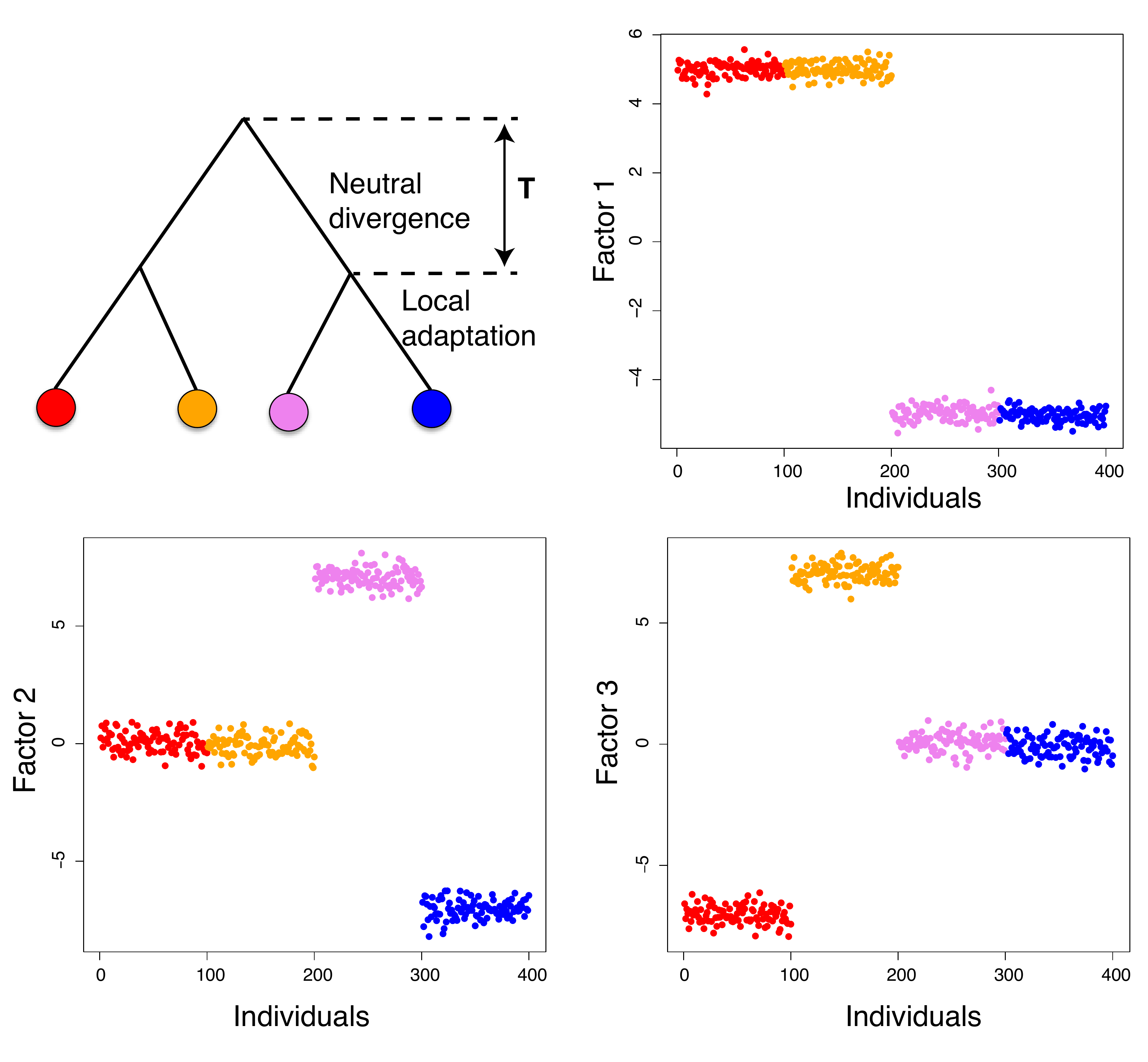}
	\caption{Values of the $K=3$ factors for a population divergence model with 4 populations. The upper left panel shows the models of population divergence. The other panels show the values of the first three factors and each dot corresponds to one individual. As candidates for local adaptation, the factor model with $K=3$ looks for SNPs whose variation is atypically well explained by one of the three factors.}
 \label{fig:split} 
\end{figure}

To provide a concrete example of how factors represent population structure, we consider a model of population divergence. We assume that an initial population splits into two populations $A$ and $B$ that diverged according to neutral evolution. The initial neutral divergence of duration $T$ is followed by 2 concomitant splits where each daughter population $A$ and $B$ splits into two subpopulations $(A_1,A_2)$ and  $(B_1,B_2)$. By contrast to the initial divergence that is purely neutral, the second phase of divergence between populations assumes some local adaptation with a small proportion of SNPs  conferring selective advantage (Figure \ref{fig:split}). We fit the factor models with $K=3$ and we display the three factors in Figure \ref{fig:split}. The first factor discriminates individuals according to the initial split and  the second and third factor discriminate individuals according to the subsequent splits which separate subpopulation $A_1$ from $A_2$ (second factor) and subpopulation $B_1$ from $B_2$ (third factor).

We now specify how we measure in a Bayesian fashion the degree of outlyingness for each locus. To account for outlier and non-outlier loci, we assume that, at a given locus $\ell$, the vector of regression coefficients ${\bf V}_{\ell}=(V_{1\ell},\dots,V_{K\ell})$ comes from a mixture of two different distributions. We introduce a vector ${\bf z}$ of indicator variables ($z_1,\dots,z_p)$ whose elements are equal to 0 for non-outlier loci and take values in ${1,\dots,K}$ for outlier loci. 
For both non outlier and outlier locus $\ell$, we assume that the vector ${\bf V}_{\ell}=(V_{1\ell},\dots,V_{K\ell})$ is composed of independent Gaussian random variables. The model for non-outlier loci is a product of Gaussian distributions
\begin{equation}
\label{eq:H0}
V_{k\ell} | z_\ell=0 \sim {\mathcal N}(0,\sigma_k^2),\; k=1,\dots,K,
\end{equation}
where ${\mathcal N}(m,\sigma^2)$  denotes the Gaussian distribution of mean $m$ and variance $\sigma^2$. To model outlier loci, we consider a variance-inflation model which assumes an inflated variance to account for outlier loci \cite[]{box68,devlin1999}. The model for outlier loci is itself a mixture model with $K$ components of equal weights where the $k^{\rm th}$ component assumes an inflated variance for the $k^{\rm th}$ regression coefficient but not for the other ones. Denoting by $c_{k}^2$ the variance-inflation parameter for factor $k$ ($c_{k}^2>1$),the $k_0^{\rm th}$ component of the mixture model for outlier assumes a product of Gaussian distributions
\begin{eqnarray}
\label{eq:H1}
V_{k\ell} | z_\ell=k  & \sim& {\mathcal N}(0,c_{k}^2 \sigma_{k}^2) ,\; k=k_0 \nonumber \\
V_{k\ell} | z_\ell=k & \sim &{\mathcal N}(0,\sigma_{k}^2), k\neq k_0
\end{eqnarray}
The model for outliers has been chosen for sake of interpretability. Each outlier locus can be related to one of the $K$ factors because outlier loci should be atypically explained by one of the $K$ factors. To measure for each locus the strength of evidence for outlyingness, we compute the Bayes factor of the outlier model against the non-outlier model. If a locus is considered as an outlier, the factor with which there is an atypical correlation is found by computing the posterior probabilities of each of the $K$ components of the outlier mixture model. To account for Linkage Disequilibrium (LD), we additionally consider a Potts Model that encourages outlier loci to be clustered in the genome (see METHODS for details). 

When fitting the factor model with $K=3$ to data simulated under the scenario of population divergence depicted in Figure \ref{fig:split}, the outlier  model of equation (\ref{eq:H1}) assumes 3 different types of outlier loci : loci that have large genetic differentiation when comparing the pair of subpopulations $(A_1,A_2)$ to the pair $(B_1,B_2)$ (large values of $|V_{1\ell}|$),  loci that  have large genetic differentiation when comparing subpopulation $A_1$ to $A_2$ (large values of $|V_{2\ell}|$), and loci that  have large genetic differentiation when comparing subpopulation $B_1$ to $B_2$  (large values of $|V_{3\ell}|$). Because the simulation assumes that the initial period of divergence is purely neutral, the first types of outliers (large values of $|V_{1\ell}|$) are in fact false positives.

\section*{Results}
\subsection*{Simulation study}
\subsubsection*{Population divergence model}

The first simulation study investigates to what extent factor models better account for population structure than methods based on $F$-statistics. We consider the model of population divergence depicted in Figure \ref{fig:split}. An initial neutral divergence is followed by adaptive divergence where $4\%$ of the 10,000 simulated SNPs are involved in local adaption. The set of adaptive SNPs  is split in four equal parts and each subset of SNP confers a selective advantage in only one of the four populations. When the initial neutral divergence time $T$ is null, the population tree is star-like, and the assumption of the $F$-model is valid. As the initial neutral divergence time $T$ increases, the departure to the $F$-model increases. The neutral divergence time $T$ is scaled so that $T=1$ means that the neutral and adaptive phases are of same duration.

\begin{figure}[ht!]
	\centering
		\includegraphics[width=16cm]{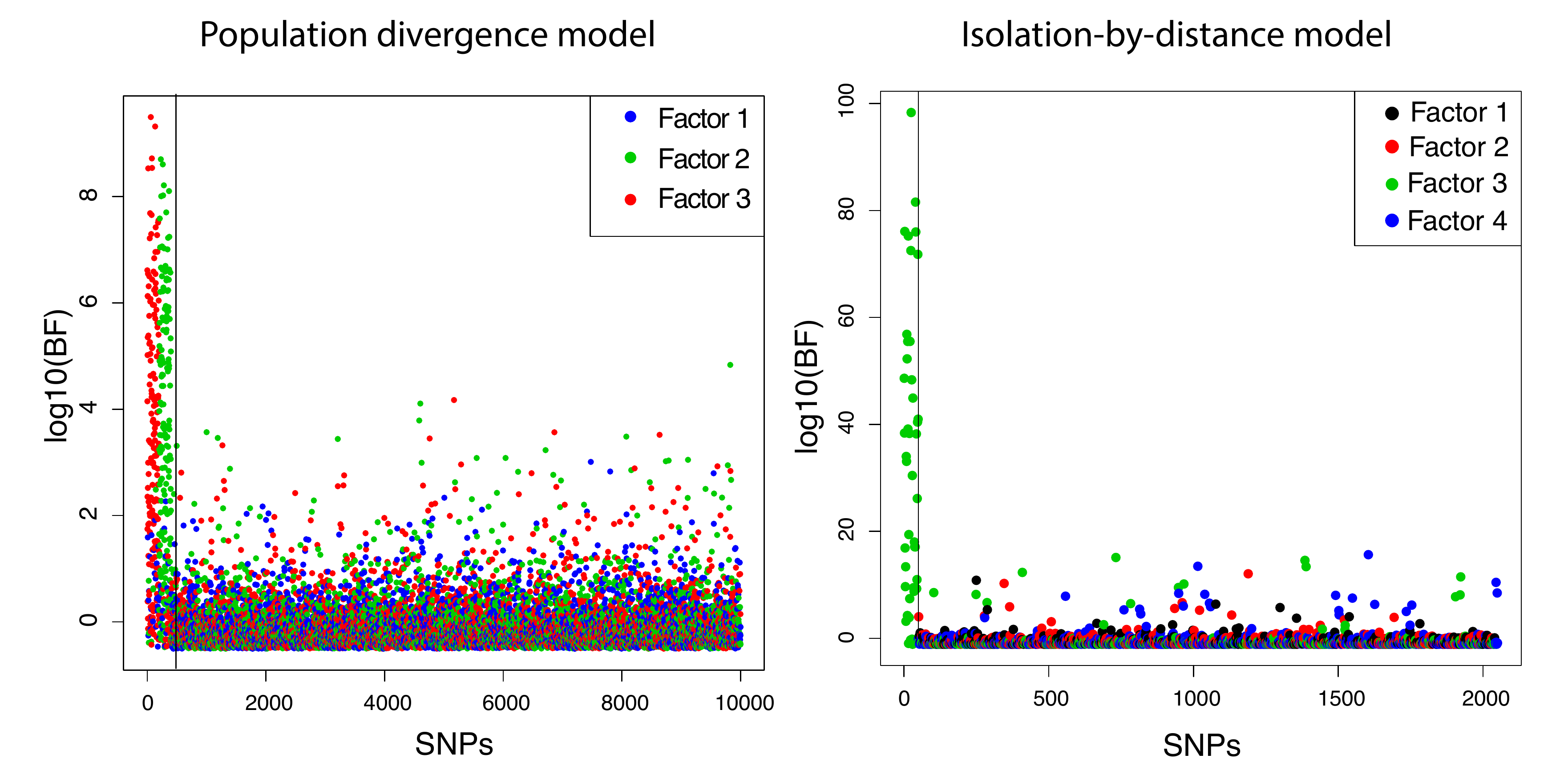}
	\caption{Bayes factors for the two different examples. The SNPs under selection are located on the left-hand side of the vertical bar.}
 \label{fig:BF} 
\end{figure}

First, we present results using the factor model with $K=3$ factors that is optimal because there are 4 populations in the divergence model \cite[]{patterson06}. We consider a long-enough divergence time $T=2$ so that the first factor corresponds to the initial and neutral divergence whereas the second and third factors correspond to the subsequent divergence events during which biological adaptation took place (Figure \ref{fig:split}). The SNPs that have been truly involved in biological adaptation are usually associated with the correct factor because among the 400 truly adaptive SNPs, $81\%$ are associated with the second and third factor and this proportion raises to $98\%$ (resp.  $92\%$) when considering the $195$ (resp. 305) adaptive SNPs with Bayes factors larger than 10 (resp. 1) (Figure \ref{fig:BF}).  
 
\begin{figure}[ht!]
	\centering
		\includegraphics[width=14cm]{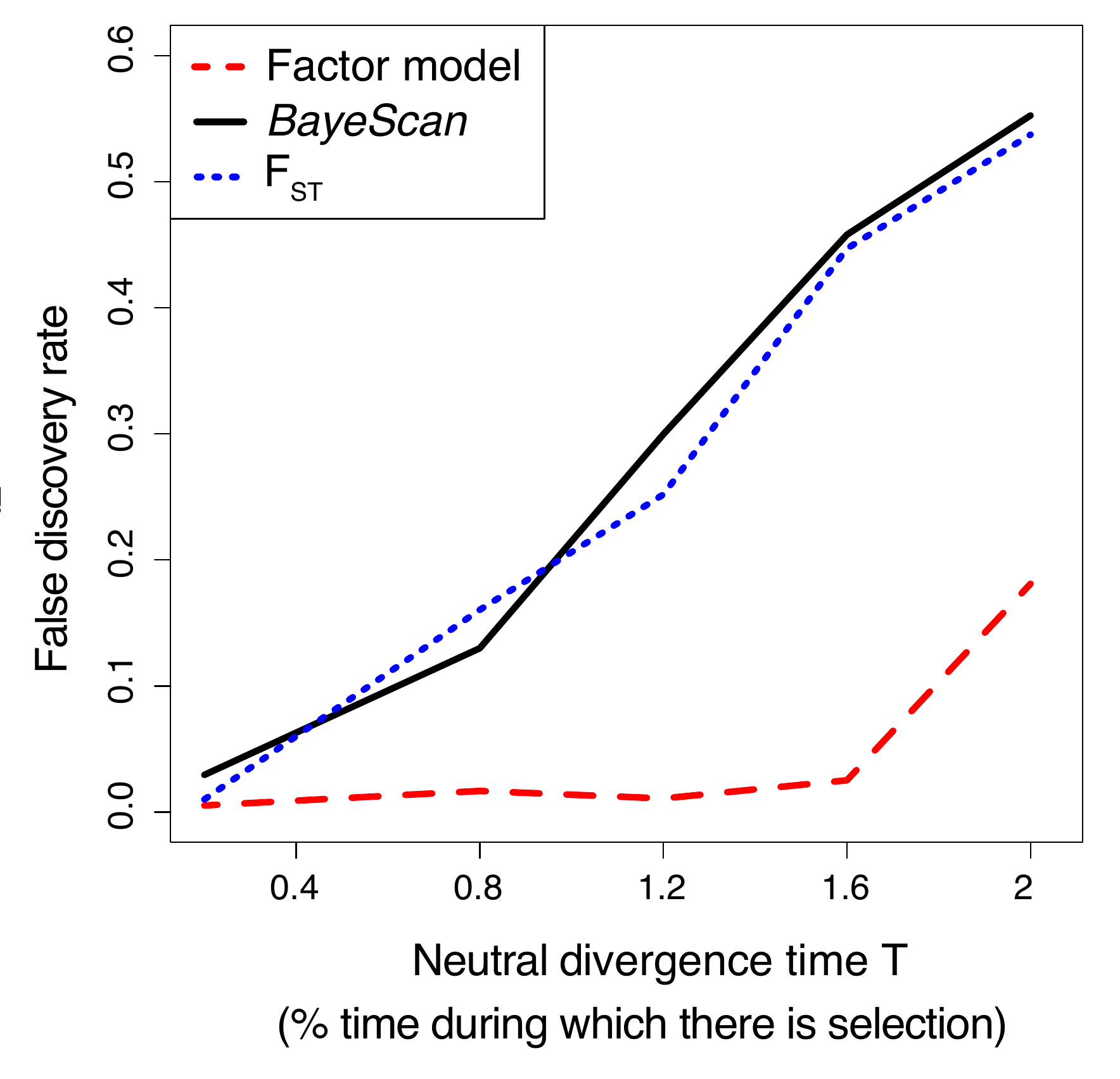}
	\caption{False discovery rate as a function of the initial divergence time $T$ in the population divergence model of Figure \ref{fig:split}. For both {\it BayeScan} and the proposed factor model, Bayes factors are used for ranking SNPs whereas we use $F_{ST}$ values for the standard genome scan based on $F_{ST}$ values. To determine a threshold above which SNPs are considered as outlier, we constrain the lists of SNPs provided by each method to contain $50\%$ of the 400 SNPs truly involved in local adaptation. The neutral divergence time $T$ is scaled so that $T=1$ means that the neutral and adaptive phases are of same duration.}
 \label{fig:fdr_mod1} 
\end{figure}

Then, we compare the false discovery rates of three different approaches including the proposed factor model, {\it BayeScan} (version 2.1), and genome scans based on the $F_{ST}$ statistic. For both {\it BayeScan} and the proposed factor model, we use Bayes factors for ranking SNPs whereas we use $F_{ST}$ values for the last method. More precisely, we use the q-values for ranking SNPs with {\it BayeScan} but by definition of the q-values, it provides the same ranking as the Bayes factors. To determine a threshold above which a SNP is considered as an outlier, we enlarge the lists of top ranked SNPs, provided by each method, until each of them contains $50\%$ of the 400 truly adaptive SNPs. This procedure amounts at setting the sensitivity to $50\%$ (the sensitivity is also called recall rate in machine learning). Figure \ref{fig:fdr_mod1} shows that for all methods, the false discovery rate (FDR) is below $5\%$ when the population tree is almost star-like ($T=0.04$) but it increases with the initial neutral divergence time $T$. Although the FDR always increases with $T$, the FDRs of the factor model are always smaller than FDRs obtained with $F_{ST}$ and with {\it BayeScan}. For instance, when $T=1$, the FDRs obtained with $F_{ST}$ and {\it BayeScan} are between $20\%$ and $30\%$ whereas it is smaller than $5\%$ with the factor model. Instead of using a threshold of $50\%$, we also constrain the lists of SNPs to contain $25\%$ or $75\%$ of the truly adaptive SNP (i.e. setting the sensitivity to  $25\%$ or $75\%$). As for the $50\%$ threshold, all methods have small FDR for small-enough initial divergence time $T$, and, as $T$ increases, FDR increases at a slower rate for the factor model (Figure S1). In summary, the false discovery rate increases as the model of divergence deviates from a star-like phylogeny, but compared to other methods, the factor model reduces the proportion of false discoveries by a factor of 2 or more when there is a strong-enough deviation from the star-like assumption ($T>0.8$).

The results presented so far were obtained using the factor model with  $K=3$ factors. By increasing the values of $K$ from 1 to 6, we find that, compared to $K=3$, the false discovery rate drastically increases for underspecification of $K$ ($K<3$) but is almost insensitive to overspecification of $K$ ($K>3$, Figure S2). We also compute the mean square error (MSE) of equation (\ref{eq:factor}) for different values of $K$ to determine if the MSE can be a guide for choosing $K$. The MSE decreases from $K=1$ to $K=3$ before staying almost constant as $K$ continues to grow (Figure S3). In this example of population divergence, the MSE suggests to choose $K=3$ but choosing a more complex model ($K>3$) would provide comparable false discovery rates.

\subsubsection*{Isolation-by-distance model}
The second simulation study provides an example of how to search for biological adaptation when there is isolation-by-distance. Approaches based on $F$-statistics would require to group individuals into populations, and we want to avoid that. On a two-dimensional 10 $\times$ 10 grid, we simulate a stepping-stone model  with selection acting on individuals located in the lower-right corner of the grid. We sample 10 diploid individuals at each of the 100 demes. A total of $50$ out of $2,050$ SNPs confer selective advantage in this region and the selection coefficient decreases gradually when moving away from the point where selection is maximal. 

\begin{figure}[ht!]
	\centering
		\includegraphics[width=14cm]{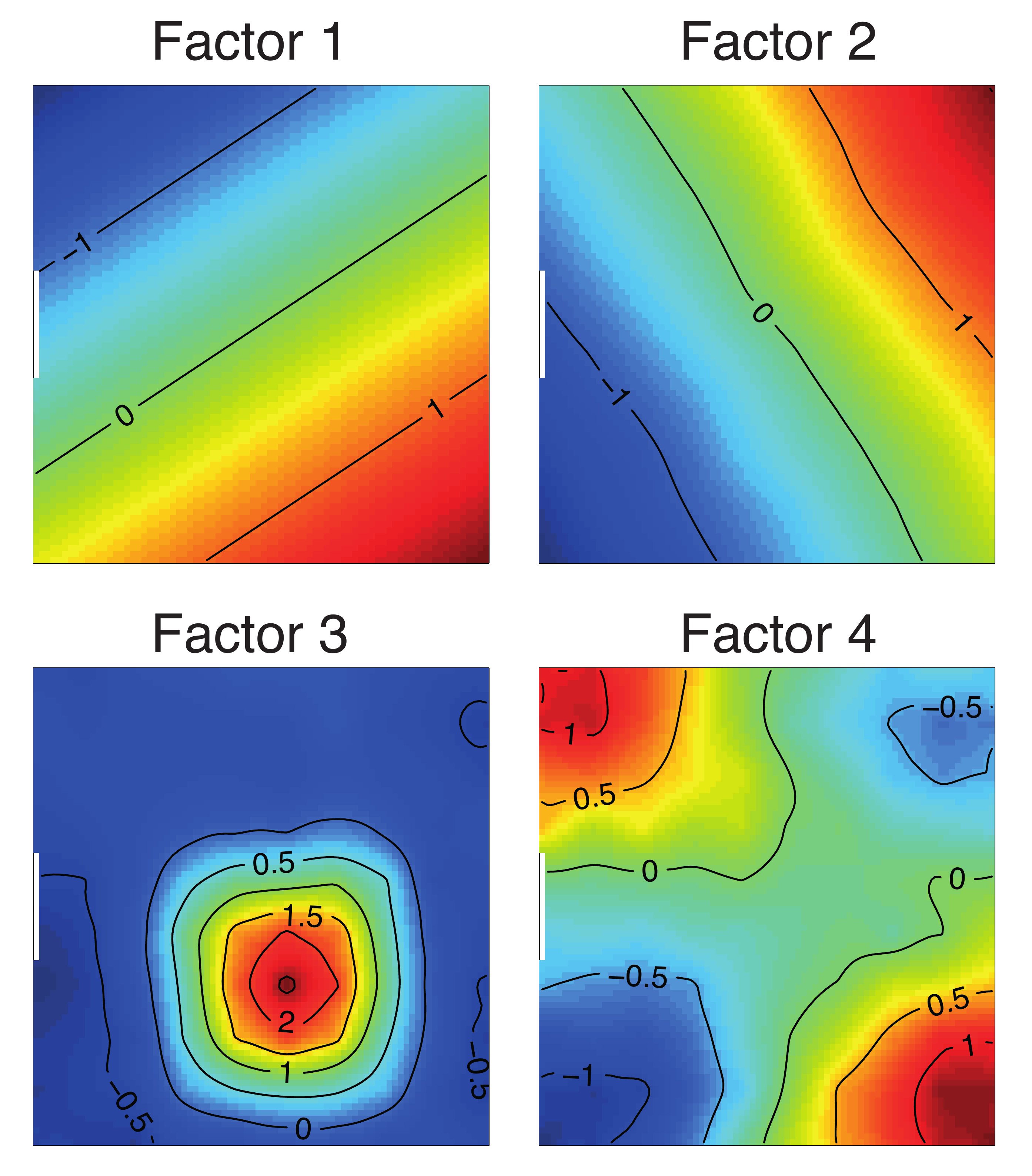}
	\caption{Values of the $K=4$ factors for the two dimensional isolation-by-distance model. The different panels shows the values of the first four factors when projected onto the two-dimensional space. As candidates for local adaptation, the factor model with $K=4$ looks for SNPs whose variation is atypically well explained by one of the four factors. Spatial interpolation of the factors is obtained using the {\it Krig} function that is available from the {\it fields}  R package.}
 \label{fig:IBD} 
\end{figure}

With the factor model, the selection gradient is reflected in a different factor depending on the value of the selection coefficients (results not shown). Here we choose the intensity of selection such as  the selection gradient is visible in the third factor (Figure \ref{fig:IBD}). The other factors have spatial patterns that are typical of isolation-by-distance models \cite[]{novembre08}. We choose $K=4$ because the MSE decreases from $K=1$ to $K=4$ before being almost constant (Figure S3). In terms of false discovery rate, this choice of $K$ is not optimal because $K=3$ would provide smaller FDR (Figure S4). However, as in the first example, smaller values of $K$ compared to the optimal value ($K<3$) increases FDR drastically whereas too large values of $K$ ($K>3$) increases the optimal FDR more moderately (Figure S4). With $K=4$, the FDR is of $0\%$ when considering the top $25$ SNPs, which corresponds to a a sensitivity of $50\%$. However, when setting the sensitivity at $75\%$, the FDR increases to $30\%$, which corresponds to $38$ true positive SNPs among a list of 54 SNPs. The 50 truly adaptive SNPs are all correctly associated with the factor corresponding to biological adaptation, which is the third factor here (Figure \ref{fig:BF}). When decreasing the number of sampled individuals from 10 to 1, the false discovery rate, obtained with a sensitivity of $50\%$, increases considerably from $0\%$ to $91\%$ (Figure S5).

\subsection*{Analysis of human SNP data}
\begin{figure}[ht!]
	\centering
		\includegraphics[width=14cm]{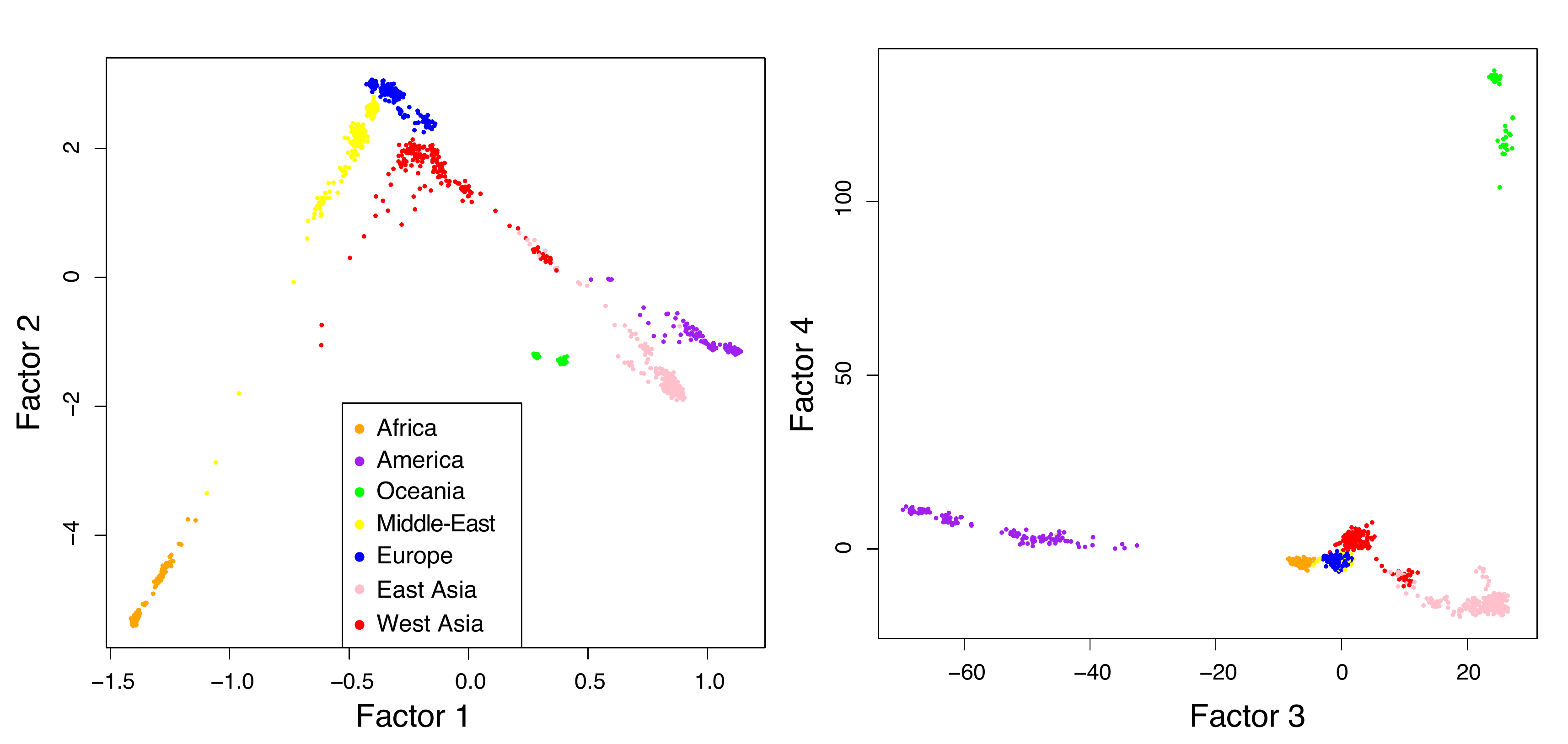}
	\caption{Values of the $K=4$ factors for the HGDP dataset.}
 \label{fig:HGDP} 
\end{figure}

The HGDP dataset contains $644,199$ SNPs, after removal of the SNPs on the sex chromosomes and on the mitochondrion, which have been typed  for $1,043$ individuals coming from $53$ different populations  \cite[]{li08}. First, we fit the factor model with different values of $K$. By contrast with the two previous examples, there is no value of $K$ at which the MSE stops to decrease (Figure S3). By looking at the different factors (Figure \ref{fig:HGDP} and Figure S6),  we decide to consider a model that captures genetic differentiation between but not within continents. Using this criterion, we consider a factor model with $K=4$ since larger values of $K$ would reveal genetic difference within continents (Figure S6). The first factor mostly contrasts African from Asiatic and Native American individuals and the second factor mainly discriminates African from Middle-Eastern and Western Eurasian individuals. The third factor distinguishes Native American individuals---coming from Central and South America---from the rest of the sample whereas the last factor separates individuals from Oceania from the rest of the sample (Figure \ref{fig:HGDP}).

We choose to restrict our analysis to the $5,000$ top-hit SNPs (Table S1). Their values of the Bayes factors range from $1.03$ to $5.05$ on a $\log_{10}$ scale. The two SNPs with the largest Bayes factors (rs1834640 and  rs2250072) are correlated with the second factor. They are located on chromosome 15 and the closest gene is {\it SLC24A5}, which is located at $20-30$ kb from the SNPs. Among the $5,000$ SNPs with largest Bayes factors, $851$ are related to factor 1, $844$ with factor 2, $1,982$ with factor 3 and $1,323$ with factor 4. For each of the four sublists, we further provide information for the 10 SNPs with the largest Bayes factors (Table \ref{tab:outliers}). 
\begin{itemize}

\item For the first factor, although we consider ten different SNPs, only two genomic regions are found. One of the two genomic regions is located on chromosome 10 and downstream of the oncogene {\it CYP26A1} whose  expression is enhanced in sunlight-damaged human skin \cite[]{osanai2011,mallick13}. The other SNPs were found in the SM6 gene which is implicated in the structural maintenance of chromosome protein 6 and which has already been picked as a candidate for selection in another scan with the HGDP sample  \cite[]{hao13}. For all the ten SNPs, we investigate the worldwide repartition of allele frequencies with the ALFRED database \cite[]{rajeevan2012}.  East Asiatic and Native American populations have allele frequencies that are different from the rest of the sample (Table S2) as can be predicted when  looking at the geographic repartition of the first factor (Figure \ref{fig:HGDP}). 

\item For the outlier SNPs associated with the second factor, the allelic frequencies were mostly different when comparing Western Eurasian individuals to the rest of the sample (Table S2). In addition to the SNPs close to the {\it SLC24A5} gene that is associated with light skin in Western Eurasia  \cite[]{canfield13}, we also find four other regions locating close to the following genes: {\it EDAR} in chromosome 2 which has been associated with various traits including hair thickness and sweating \cite[]{kamberov13}, {\it SLC35F3} in chromosome 1, {\it KIF3A} in chromosome 5, {\it RABGAP1} and {\it STRBP} in chromosome 9 with the latter being involved in spermatogenesis, and {\it MYO5C} and {\it DUT} in chromosome 15.

\item For the third factor, eight out of the ten SNPs with the largest Bayes factor are found in a 1Mb region of chromosome 22, which encompasses many different genes (Table \ref{tab:outliers}). For the SNPs in this large region of chromosome 22, the allelic frequencies mostly differ between Native Americans and the rest of the sample. For Sub-Saharan African populations, allelic frequencies of these SNPs are intermediate with Pygmies populations having frequencies that are often the most similar to the Native Americans (Table S2).

\item The allele frequencies of the SNPs that are the most associated with the fourth factor mostly differ between individuals from Oceania (Papuan and Melanesian) and the rest of the sample with Native Americans and Pygmies population having, for some SNPs, allele frequencies  that are the most similar to the Oceanians (Table S2). Among the ten outlier SNPs, four SNPs are located in chromosome 8 and four SNPs are located in chromosome 17. Among the $1,323$ SNPs associated with the fourth factor, there is an excess of outlier SNPs in chromosome 8 (Figure S7) pointing to a prominent role of its genes in adaptation to the local conditions of Oceania. There are different genomic regions with large Bayes factors in chromosome 8 and one of these genomic regions encompasses {\it RP1L1}, a gene often found in selection scan \cite[]{barreiro08} and related to eye diseases \cite[]{davidson13}.
\end{itemize}

We also perform a Gene Ontology (GO) enrichment analysis on human genes using the 5,000 SNPS with the largest Bayes factors. We find significant enrichment of biological processes for each of the four factors (Table S3). Some interesting instances of the enriched gene ontologies include three different GO terms related to regulation of hormone secretion for the first factor, enrichment of homophilic cell adhesion for the third factor, and aging for the fourth factor. Finally, we look at a catalog of published GWAS \cite[]{welter14} to search for enrichment of outlier SNPs related to a particular phenotype (Table S4). The traits that are the most associated with the outlier SNPs are height (6 SNPs), obesity and weight (5 SNPs) and Crohn's disease (5 SNPs).

\section*{Discussion}

Based on a Bayesian factor model, we provide a new approach for performing genome scan of local adaption. The hierarchical factor model considers as outliers the SNPs that are atypically related to population structure. A set of $K$ latent factors measure population structure and they can adequately describe clustering of individuals into populations (Figure \ref{fig:split}), isolation-by-distance patterns and gradients of selection (Figure \ref{fig:IBD}). Compared to the software {\it BayeScan} or to genome scans based on $F_{ST}$, the factor model does not assume a particular model of population structure. In a model of population divergence, we show that removing the assumptions of the $F$-model considerably reduces the false discovery rate. To explain why the factor model generates less false discoveries, we introduce the notions of mechanistic and phenomenological models \cite[]{hilborn97}.  Mechanistic models aim to mimic the biological processes that are thought to have given rise to the data whereas phenomenological models seek only to best describe the data using a statistical model. In the spectrum between mechanistic and phenomenological model, the $F$-model would stand close to mechanistic models whereas factor models would be closer to the phenomenological ones. Mechanistic models are appealing because they provide quantitative measures that can be related to biologically meaningful parameters. For instance, $F$-statistics measure genetic drift which can be related to migration rates, divergence times or population sizes. By contrast, phenomenological models work with mathematical abstractions such as latent factors that can be difficult to interpret biologically. The downside of mechanistic models is that violation of the modeling assumption can invalidate the proposed framework and generate many false discoveries in the context of selection scan. The $F$-model assumes a particular covariance matrix between populations which is found with star-like population trees for instance \cite[]{bonhomme10}. However, more complex models of population structure can arise for various reasons including non-instantaneous divergence and isolation-by-distance, and they will violate the mechanistic assumptions and make phenomenological models preferable.

Although principal component analysis or the related factor model are generally used to investigate population structure, there have been already several attempts at performing selection scan based on these statistical approaches. A first idea is to compute $F_{ST}$ values between pairs of populations that contain the top and bottom individuals for each principal component \cite[]{abdellaoui13}. This approach provides a list of outliers that are specific to each principal component in the same way as the hierarchical factor model of equations (\ref{eq:factor})-(\ref{eq:H1}) provides outliers that are related to one of the $K$ factors. A second proposition involves new interpretations of PCA  related to $F$-statistics, which provide statistical measures to detect local adaptation \cite[]{laloe11}. A last and recent proposition called ``logistic factor analysis''\ adds a logistic link function to the factor model (equation (\ref{eq:factor})) in order to guarantee that the predicted values can be interpreted as frequencies because they lie between 0 and 1 \cite[]{hao13}. Loci involved in biological adaptation were scanned using a deviance statistic \cite[]{hao13}. These related approaches are built on the success of PCA and factor models to capture population structure with a small number of variables.

Choosing the dimension $K$ of the statistical model that ascertains population structure is a recurrent problem. One possibility is to use an {\it objective} approach based on a quantitative criterion. Examples of such objective criteria include the $\Delta_K$ measure to detect the number of clusters using the software STRUCTURE \cite[]{evanno05} or the Tracy-Widom statistic to choose the dimension $K$ in PCA \cite[]{patterson06}. Another possibility is to adopt a {\it subjective} approach and to choose a value of $K$ such that increasing $K$ would provide results that are considered of too little interest. With the proposed Bayesian factor model, we implemented both approaches. For the simulations, choosing $K$ based on the mean square error (MSE) of equation (\ref{eq:factor}) works well because the MSE stops to decrease when $K$ increases beyond a certain value. However, for the human data, the choice is more complex because the MSE decreases regularly as $K$ increases. We chose $K=4$ because we were only interested in biological adaptation that is related to genetic differentiation between continents but we acknowledge that major adaptive processes although occurs within continents \cite[]{jarvis2012}. To provide recommendations for choosing $K$, we suggest to fit the hierarchical factor model with different values of $K$ in order to investigate if there is a value of $K$, at which the MSE stops or almost stops to decrease. If not, the choice of $K$ can be based on subjective arguments where the latent factors of too little interest can be discarded. 

One of our objective was to propose a method for selection scan that avoids the computational burden of some Bayesian approaches, which can become a serious obstacle when analyzing large scale SNP data. This objective is however only partly fulfilled. The downside of our approach is that it relies on a MCMC algorithm that quickly grinds to a
halt under the sheer mass of SNP data \cite[]{lange2014}. Hopefully, the MCMC algorithm is based on a Gibbs sampler that alternates the  computation of least square solutions, which are fast to compute. For the HGDP dataset ($644,199$ SNPs), the run-time ranges from $13$ to $16$ hours using a single computer processor (2.4 GHz 64bit Intel Xeon) when $K$ increases from 1 to 8. Because dataset containing millions of SNPs are becoming available, we are currently working on the development of  a faster version of our software.

Fitting the factor model to the HGDP data, we kept the $5,000$ SNPs with the largest Bayes factors. The first two factors mainly measure differentiation  between Africa, Western Eurasia and East Asia and some outliers related with these two factors are involved in morphological traits, which is a trait often reported to be enriched with genes having signatures of positive selection \cite[]{barreiro08}. In the list of outliers, we found the genes  {\it SLC24A5} and {\it EDAR}, often reported as top hits in selection scan \cite[]{pickrell09,hao13}, and related to skin pigmentation and hair thickness respectively \cite[]{kamberov13,mallick13}. We also found that SNPs close to the oncogene {\it CYP26A1} whose expression is enhanced in sunlight-damaged human skin \cite[]{osanai2011} is part of the top list for outliers. The third and fourth factors correspond to genetic differentiation between Native Americans, individuals from Oceania and the rest of the sample. There are many regions in chromosome 8 enriched with outlier SNPs. One of this region encompasses the gene {\it RP1L1} that is associated with retinal diseases and which has already reported to have one of the strongest signature of positive selection along with other genes related to sensory functions \cite[]{barreiro08}. Many outlier SNPs strongly related to the third and fourth factors have allele frequencies that are similar between Southern Native Americans and Pygmies (third factor) or between individuals from Oceania, Southern Native Americans and Pygmies (fourth factor). Because these individuals all live in tropical rain forests and have similar diet consisting of roots and tubers, our findings support the importance of diet, climate, and potentially pathogen load to explain human adaptation \cite[]{hancock2010,fumagalli11}. The SNPs with similar allele frequencies in different geographic regions are good candidates for convergent evolution and  would deserve further analysis. 

Factor models are enriching the toolbox of population genetic methods. The main principle is to model population structure via latent variables called factors. Factors models have already been proposed to ascertain population structure \cite[]{engelhardt10}, and to  account for population structure when testing for gene-environment association \cite[]{frichot13}. We showed that factor models also provide a convenient individual-based framework to find loci that have atypical patterns of genetic differentiation. A major argument supporting the proposed hierarchical factor model is that it produces less false discoveries compared to genome scans based on $F_{ST}$.

\section*{Materials and Methods}

\subsection*{Hierarchical Bayesian modeling}
We provide the prior distributions for the latent variables of the hierarchical factor model defined by equations (\ref{eq:factor})-(\ref{eq:H1}).
To account for Linkage Disequilibrium (LD) in the genome and encourage outlier loci to be clustered along the genome, we consider a Potts model with external field for the indicator variables \cite[]{winkler03}
\begin{equation}
\label{eq:potts}
p(z_1,\dots,z_p) \propto  (1-\pi)^{p_0}\pi^{(p-p_0)} e^{\beta \sum_{i \sim j} \mathbbm{1}_{z_i=z_j}},
\end{equation}
where the sum in the exponential ranges over all pairs of neighboring loci. We consider that each locus has two neighbors except at the beginning and at the end of the chromosome where a locus has only one neighbor. In equation (\ref{eq:potts}), the variable $p_0$ is the number of loci such that $z_i=0$, $\mathbbm{1}$ is the indicator function, $\beta$ is the parameter of the Potts model and is set to $\beta=1$, and $\pi$ is the prior proportion of outlier. To model the proportion of outlier, we consider a uniform prior on the $\log_{10}$ scale reflecting that we are interested in the order of magnitude of the proportion of outlier loci \cite[]{guan11}. In the following we consider $-4$ and $-1$ for the lower and upper bound of the uniform prior. For the variance parameters $\sigma_k^2$, $k=1,\dots,K$, that are specific to each factor (equations (\ref{eq:H0}) and (\ref{eq:H1})), we consider the parametrization $\sigma_k^2=\sigma^2 \rho^2_k$ where $\sigma^2$ is the residual variance in equation (\ref{eq:factor}) \cite[]{oba03}. We consider the non-informative prior for the variance parameters $p(\sigma^2) \propto 1/\sigma^2$ and $p(\rho^2_k) \propto 1/\rho^2_k$, $k=1,\dots,K$. For the variance-inflation parameters $c^2_k$ of equation (\ref{eq:H1}), we consider uniform priors with $1$ and $10$ for the lower and upper bounds.

With the factor model of equation (\ref{eq:factor}), there is a well known issue of identifiability because identical likelihood values can be obtained from a solution $(\bf{U},\bf{V})$ after using orthogonal rotations \cite[]{west03}. To add constraints to the model, we consider a prior with unit variance for each of the factors
$$
{\bf U}_k \sim \mathcal{N}(0,\bf{I_n}),
$$
where $\bf{I_n}$ is the squared $n\times n$ identity matrix \cite[]{oba03}. To further prevent the MCMC algorithm to produce alternative rotations of the factors \cite[]{engelhardt10}, we consider the solution of the singular value decomposition as starting values for the factors $\bf{U}_1$,\dots,$\bf{U}_K$ in the MCMC algorithm.

To evaluate the strength of evidence for outlyingness at each locus, we compute the Bayes factor on a $\log_{10}$ scale. The Bayes factor is defined as the ratio between the posterior odds  $p(z_\ell>0 | Y)/p( z_\ell=0 | Y)$ and the prior odds $p(z_\ell>0)/p( z_\ell=0)=\pi/(1-\pi)$. The description of the MCMC algorithm and of the computation of the Bayes factor is given in the supplementary material.

\subsection*{Simulation of the four-population divergence model}

The first simulation scenario is a divergence model with 4 populations. Populations have a constant effective population sizes of $N_e = 1,000$ diploid individuals, with $50$ individuals sampled in each population. The genotypes consist of $10,000$ independent SNPs. 
The simulations are performed in two steps. In the first step, we use the software {\it ms} to simulate a neutral divergence model \cite[]{hudson2002}. When looking backward in time, we instantly merge population $A_1$ with $A_2$, and population $B_1$ with $B_2$, then after waiting a number $T=20, 80, 120 ,160 ,200$ of generations, we merge the two remaining populations A and B. We keep only variants with a minor allele frequency larger than $5\%$ at the end of this first step. The second step is performed with the software {\it SimuPOP} \cite[]{peng05}. To run  {\it SimuPOP}, we provide the allele frequencies in each of the 2 populations that have been generated with {\it ms}. Looking forward-in-time, we simulate $100$ generations after the 2 concomitant divergence events. We assume no migration between populations. In each evolutionary lineage, we assume that  $100$ SNPs confer selective advantages using a selection coefficient of  $s = .1$ for homozygotes carrying two adaptive alleles. In both simulation schemes, we assume an additive model for selection. 

\subsection*{Simulation of the stepping-stone model}
The second simulation scenario is a 2-dimensional stepping-stone model with a $10\times10$ grid. Each of the 100 populations has an effective population size of $N_e = 1,000$ diploid individuals. We sample $10$ individuals in each population, and there are $2,050$ independent SNPs. 
We also consider a  2-step procedure for the simulations. First, we simulate an equilibrium  stepping-stone model with  the software {\it ms}. Neighboring populations exchange migrants with a rate $4N_e m = 8$ per generation. Then we superimpose a selection gradient using {\it SimuPOP}. During $100$ generations, we consider that $50$ SNPs confer selective advantage. The selection coefficient $s=0.1$ is maximal in population $64$, which is located in the lower right quarter of the grid.  In the four neighboring population, the selection coefficient is of $s=0.05$ and in the second layer of neighbors the selection coefficient is of $s=0.025$. The selection coefficient is equal to 0 for the rest of the grid.

\subsection*{Gene ontology analysis}
We perform a Gene Ontology (GO) enrichment analysis with {\it Gowinda} \cite[]{kofler2012}. The list of genes is built by considering all genes which contain outlier SNPs with a tolerance of $5,000$ base pairs upstream and downstream. We use a threshold of 0.05 for the false discovery rate, and we remove gene ontology terms that are shared by less than 10 genes or more than 1,000 genes. We consider the {\it --snp} flag in  {\it Gowinda}  that assumes independence of SNP within gene because the {\it --gene} flag, which assumes complete dependence of SNPs within a gene, does not provide any discovery with a FDR smaller than $5\%$. For each factor, we consider the ten gene ontology terms with the smallest false discovery rates, and we only report gene ontology terms that are related to biological processes.

\subsection*{Software availability} The computer program {\it PCAdapt} for fitting the factor model is available from the authors' web-sites (http://membres-timc.imag.fr/Michael.Blum/,
http://membres-timc.imag.fr/Nicolas.Duforet-Frebourg/).

\subsection*{Acknowledgments} This work has been partially supported by the LabEx PERSYVAL-Lab (ANR-11-LABX-0025-01).
\bibliographystyle{evolution}
\renewcommand\refname{Literature Cited}
\bibliography{pcadapt_bib}

\clearpage

\clearpage


\clearpage


\clearpage


\clearpage



\begin{table}
$\begin{array}{llllll}
\hline
\mbox{Chromosome}        &        \mbox{rs identifier}&        \mbox{closest gene} & \mbox{dist.$^{*}$}  & \log_{10}({\rm BF})^{+}& \mbox{Factor no.}\\
\hline
\mbox{chr 10}&	4918664&	{\it CYP26A1}	&	83&  2.8 & 1\\
\mbox{chr 10}	&10882168	&{\it CYP26A1}	&	92&  2.7& 1\\
\mbox{chr 10}&	7091054	&{\it MYOF}		&48& 2.7& 1\\
\mbox{chr 10}&	11187300&{\it CYP26A1}&		83&  2.6& 1\\
\mbox{chr 2}&	7556886	&{\it SMC6}		&0&  2.6& 1\\
\mbox{chr 10}&	12220128&{\it MYOF}&		91&  2.5& 1\\
\mbox{chr 10}&	6583859	&{\it CYP26A1}&		56&  2.5& 1\\
\mbox{chr 10}&	4918924	&{\it MYOF}	&	89&  2.4& 1\\
\mbox{chr 2}&	1834619	&{\it SMC6}	&	0&  2.4& 1\\
\mbox{chr 2}&	4578856	&{\it SMC6}	&	0&  2.4& 1\\

\hline

\mbox{chr 15}         &1834640        &{\it SLC24A5}        &21&  5.1&2\\
\mbox{chr 15}         &2250072        &{\it SLC24A5 }       &29&  4.2&2\\
\mbox{chr 2}           &260714        &{\it EDAR }       &        0&  3.2&2\\
\mbox{chr 1}           &7531501        &{\it SLC35F3}        &        0&  2.9&2\\
\mbox{chr 15}         &11637235&        {\it DUT} &        0&  2.9&2\\
\mbox{chr 9}           &10760260&        {\it RABGAP1}        &        0&  2.9&2\\
\mbox{chr 9}           &2416899        &{\it STRBP}       &        0&  2.8&2\\
\mbox{chr 5}           &2406410        &{\it KIF3A}        &0&  2.8&2\\
\mbox{chr 15}         &3751631        &{\it MYO5C}        &        0&  2.8&2\\
\mbox{chr 9}           &618746        &{\it RABGAP1}        &0&  2.8&2\\

\hline

\mbox{chr 22}	&139553	&{\it MEI1} &0& 4.1&3\\
\mbox{chr 22}	&5996039	&{\it PMM1}&0& 4.0&3\\
\mbox{chr 22}	&8139993	&{\it DESI1}&0& 4.0&3\\
\mbox{chr 22}	&126092	&{\it MEI1}&0& 4.0&3\\
\mbox{chr 22}	&1005402	&{\it XPNPEP3}&0& 3.6&3\\
\mbox{chr 22}	&8137373	&{\it ZC3H7B}&0&3.6&3\\
\mbox{chr 22}	&133074	&{\it MCHR1}&0&3.6&3\\
\mbox{chr 22}	&9611613	&{\it CSDC2} &0&3.5&3\\
\mbox{chr 20}	&2424641	&{\it SYNDIG1}& 71&3.5&3\\
\mbox{chr 14}	&2600814	&{\it LINC00871} &13&3.5&3\\

\hline

\mbox{chr 8}&	16892216	&{\it MAL2}&	19 &2.8&4\\
\mbox{chr 8}&	6990312&	{\it SYBU}	&	0& 2.8&4\\
\mbox{chr 17}&	9908046&	{\it MMD}&		64& 2.7&4\\
\mbox{chr 17}&	575873&	{\it MEOX1}	&	18& 2.6&4\\
\mbox{chr 4}&	4691075&	{\it NPY1R}	&	0& 2.6&4\\
\mbox{chr 17}&	4471745	&{\it MMD}	&	70&2.5&4\\
\mbox{chr 14}&	12891534& {\it CEP128} &		0&2.5&4\\
\mbox{chr 8}&	6988341	&{\it SYBU}	&	2&2.5&4\\
\mbox{chr 8}	&12216712&	{\it MSRA}		&16&2.5&4\\
\mbox{chr 17}&	11869714&	{\it MYCBPAP}&	0&2.5&4\\
\end{array}
$

{\small $^{*}$ dist. is the distance from  the closest gene and is measured in kilo base pairs.\\
$^{+} \log_{10}({\rm BF})$ is the logarithm (in base 10) of the Bayes factor}
\caption{List of the 10 SNPs with largest Bayes factors for each of the four factors obtained with the HGDP dataset.}
\label{tab:outliers}
\end{table}

\setcounter{figure}{0} \renewcommand{\thefigure}{S\arabic{figure}}
\setcounter{table}{0} \renewcommand{\thetable}{S\arabic{table}}

\clearpage
{\center \bf Supplementary Material of "Genome scans for detecting footprints of local adaptation using a Bayesian factor model."}

\begin{itemize}
\item Description of the MCMC algorithm and of the computation of the Bayes factors.
\item Figure S1: Population divergence model: false discovery rate as a function of the initial divergence time when the sensitivity (recall) has been set to $25\%$ and $75\%$.
\item Figure S2: Population divergence model: false discovery rate obtained with the factor model using different values of the number of factors $K$. 
\item Figure S3: Two example models and HGDP: average mean square error as a function of the number of factors $K$.
\item Figure S4: IBD model: false discovery rate as a function of the number of factors $K$.
\item Figure S5: IBD model: false discovery rate as a function of the number of sampled individuals per deme.
\item Figure S6: HGDP dataset: Plot of factors 5-8.
\item Figure S7: HGDP dataset: Number of outlier SNPs per chromosome.
\item Table S1: HGDP dataset: List of the 5,000 SNPs with largest Bayes factors (\href{http://membres-timc.imag.fr/Michael.Blum/publications/TabS1.csv}{csv file}).
\item Table S2: HGDP dataset: List of the $4\times10$ SNPs with largest Bayes factors and links to the {\it ALFRED} database to display allele frequencies.
\item Table S3: Results of the Gene Ontology based on the set of genes found in Table S1.
\item Table S4: List of GWAS SNPs that are found for the outlier SNPs (\href{http://membres-timc.imag.fr/Michael.Blum/publications/TabS4.csv}{csv file}).
\end{itemize}
\clearpage

\thispagestyle{empty}

\section*{MCMC algorithm}
We model the variance parameter of the regression coefficients using the parametrization $\sigma^2_k=\sigma^2 \rho^2_k$. In the following, we consider $\Sigma_V=\sigma^2 \times {\rm Diag}(\rho^2_1,\cdots,\rho^2_K)$ where ${\rm Diag}(\rho^2_1,\cdots,\rho^2_K)$ is a diagonal matrix with $(\rho^2_1,\cdots,\rho^2_K)$ in the diagonal. $\Sigma_{V,k0}$ is the same as $\Sigma_V$ except that the $k_0^{\rm th}$ element in the diagonal is $\sigma^2 c^2_{k_0} \rho^2_{k_0}$ instead of $\sigma^2 \rho^2_{k_0}$. For a given value of $K$, we describe below a single step of the MCMC algorithm where we use Gibbs updating steps except when updating $c_1^2,\dots,c_K^2$ and $\pi$.

\begin{itemize}
\item 
$$
{\bf V}_{\ell} \leftarrow \mathcal{N}({\bf m}_{\ell},\Sigma_{\ell}), \, \ell=1,\dots,p,
$$
with
$ 
{\bf m}_{\ell}=(\sigma^2\Sigma_{V, k_0}^{-1} + {\bf U}^{T}{\bf U})^{-1}({\bf U}{\bf Y}_{\ell})
$
and
$
\Sigma_{\ell}=(\Sigma_{V, k_0}^{-1} + \frac{1}{\sigma^{2}}{\bf U}^{T}{\bf U})^{-1}.
$

\item 
$$
\rho_{k}^2 \leftarrow IG(\frac{p}{2}, \frac{1}{2 \sigma^2}\sum_{\ell= 1}^p \frac{V_{k\ell}^2}{(c_k^2)^{\mathbbm{1}_{z_{\ell}=k}}}),
$$
where $\mathbbm{1}$ is the indicator function and IG is the inverse-gamma distribution.

\item
$$
\sigma^2 \leftarrow IG(\frac{(n+K)p}{2}, \frac{1}{2}\sum_{\ell = 1}^p (\sum_{i = 1}^n |Y_{i,\ell} - U_i V_{\ell}|^2 + \sum_{k=1}^K \frac{V_{k\ell}^2}{\rho_k^2 (c_k^2)^{\mathbbm{1}_{z_{\ell}=k}}})),
$$

\item
$$
{\bf U}_i \leftarrow \mathcal{N}({\bf m}_i,\Sigma), \, i=1,\dots,n,
$$
with 
$
{\bf m}_i=(I_K + \frac{1}{\sigma^{2 }}{\bf V}{\bf V}^{T})^{-1}\frac{1}{\sigma^{2}}{\bf V} {\bf Y}_i
$ and 
$
\Sigma=(\frac{1}{\sigma^{2 }}{\bf V}{\bf V}^{T} + I_K)^{-1}$. The ${\bf U}_i$'s are vectors of dimension $K$ and are the row vectors of the matrix ${\bf U}$. Note that in the main text, we rather refer to the column vectors of the matrix ${\bf U}$ when using the notation ${\bf U}_i$.

\item
$$
z_{\ell} =k \; \mbox{with  a probability} \; p(z_{\ell} = k | \pi, V_l, \Sigma_{V}, c_k^{2}), \; k=1,\dots,K
$$
with 
\begin{equation}
\label{eq:pz}
p(z_{\ell} = k | \pi, V_l, \Sigma_{V}, c_k^{2}) \propto \frac{\pi}{K} e^{\beta \sum_{i \sim \ell} \mathbbm{1}_{z_i,z_{\ell}}} \frac{1}{c}e^{\frac{V_{k\ell}^2}{2 \sigma^2 \rho_k^2}(1-1/c_k^2)},
\end{equation}
and $p(z_{\ell} = 0 | \pi, V_l, \Sigma_{V}, c_k^{2}) \propto (1-\pi)e^{\beta \sum_{i \sim \ell} \mathbbm{1}_{z_i,z_{\ell}}}$ for $k=0$.
\item Update $(c_1^2,\dots,c_K^2)$ using a Metropolis-Hasting step where the proposal for each element is a Gaussian random walk of variance equal to 1.

\item Update $\log_{10}(\pi)$ using a Metropolis-Hasting step where the proposal is a Gaussian random walk of variance equal to $0.5$.
\end{itemize}
We find that MCMC runs of 400 iterations (burn-in of 200 iterations) are long-enough to provide convergence.
\section*{Bayes factors}

For each SNP, the Bayes factor is defined as the ratio between the posterior odds (outlier vs non-outlier) and the prior odds $\pi/(1-\pi)$. For numerical convenience, we rather compute the logarithm (in base 10) of the Bayes factor using a Monte-Carlo approximation. The $\log_{10}$ of the posterior odds (LPO) can be computed along the MCMC run by averaging
\begin{equation}
\label{eq:lpo}
LPO=\log_{10}\{\sum_{k=1}^K p(z_{\ell} = k | \cdot)\}- \log_{10}\{p(z_{\ell} = 0 | \cdot)\},\; \ell=1,\dots,p.
\end{equation}
where $p(z_{\ell}  | \cdot)$ is the conditional probability of $z_{\ell}$ given the other parameter values. Because the maximum of the probabilities  $p(z_{\ell} = 1 | \cdot),\cdots, p(z_{\ell} = K | \cdot)$ is on a much larger scale compared to the other probabilities, we approximate equation (\ref{eq:lpo}) using
 $$
 LPO \sim \log_{10}\{\max_{k=1,\dots,K} p(z_{\ell} = k | \cdot)\}- \log_{10}\{p(z_{\ell} = 0 | \cdot)\},
 $$
and the conditional probabilities $p(z_{\ell}  | \cdot)$ are given in the equation (\ref{eq:pz}) of the supplementary material. 

\clearpage


\begin{figure}
   \includegraphics[scale=.5]{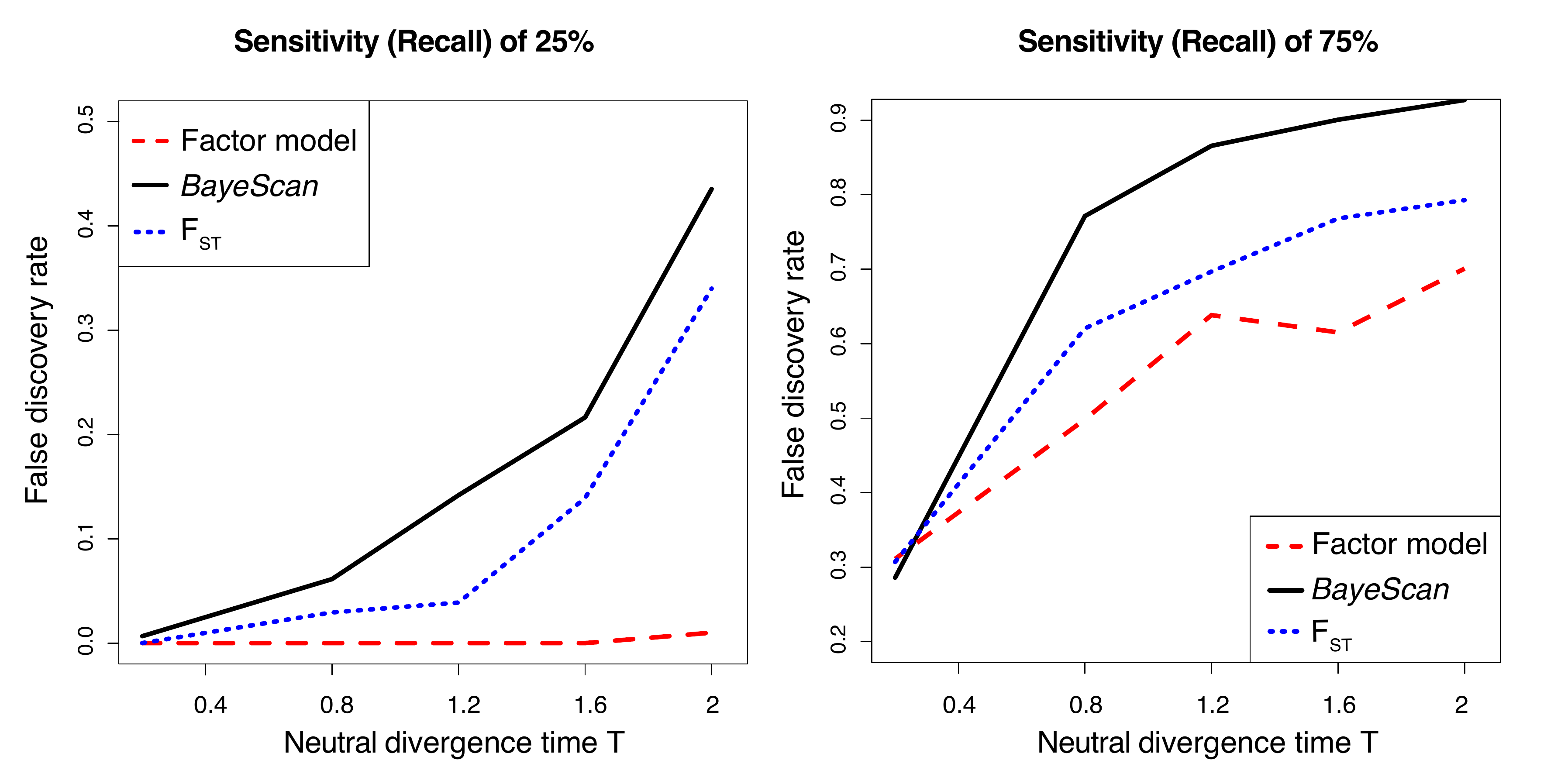}
    \caption{False discovery rate as a function of the initial divergence time in the population divergence model depicted in the Figure 1 of the main text. For both {\it BayeScan} and the proposed factor model, Bayes factors are used for ranking SNPs whereas we use $F_{ST}$ values for genome scans based on $F_{ST}$ values. To determine a threshold above which SNPs are considered as outlier, we constrain the lists of top-ranked SNPs provided by each method to contain $25\%$(left panel) or $75\%$ (right panel) of the 400 SNPs truly involved in local adaptation. Time is counted in units of time during which there is selection.}
\end{figure}

\clearpage

\begin{figure}
   \includegraphics[scale=1]{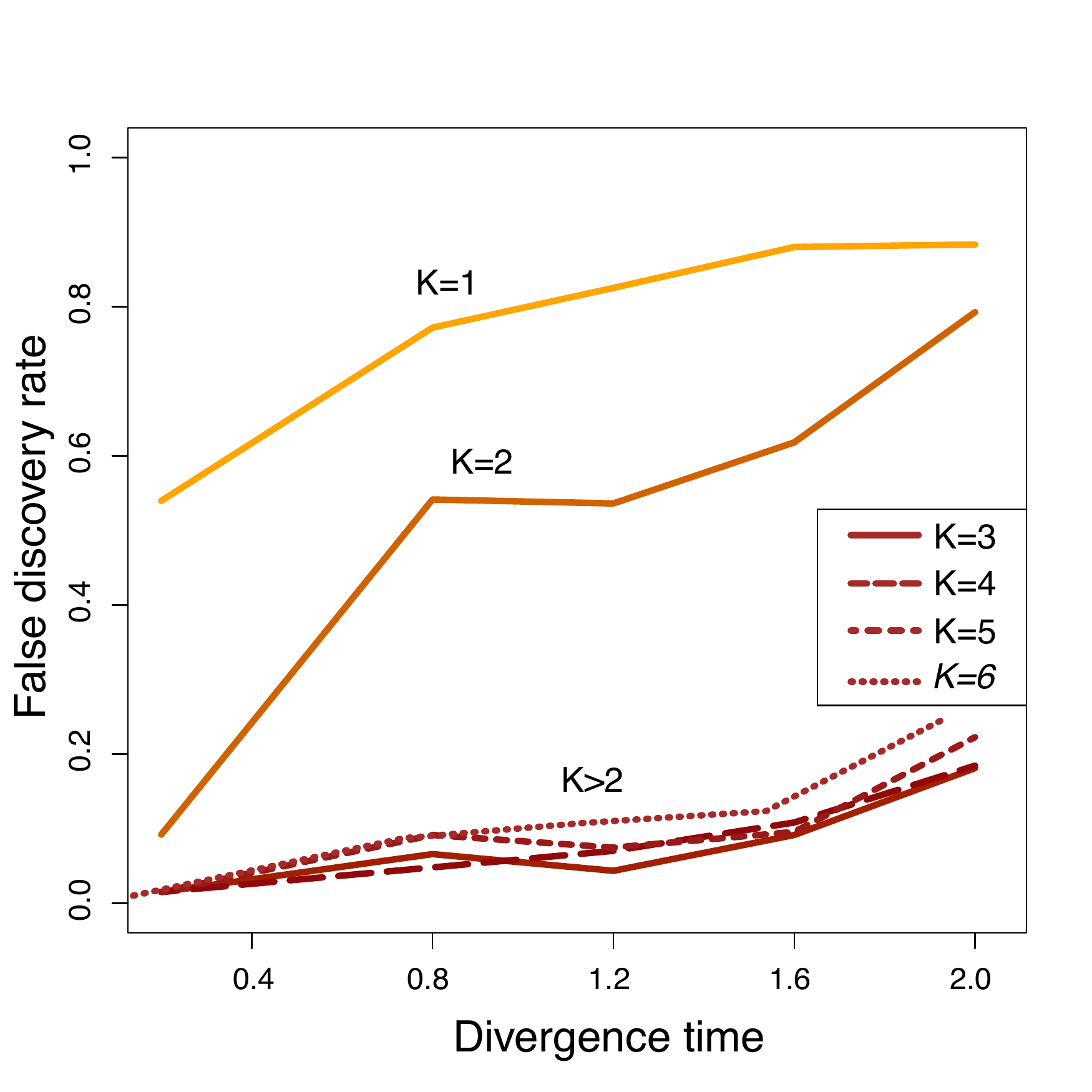}
    \caption{False discovery rate obtained with the factor model using different values of the number of factors $K$. Data are simulated with the population divergence model of the Figure 1 of the main text. To determine a threshold above which SNPs are considered as outlier, we constrain the lists of top-ranked SNPs provided by each method to contain $50\%$ of the 400 SNPs truly involved in local adaptation.}
\end{figure}
  
  \clearpage

\begin{figure}
   \includegraphics[scale=.5]{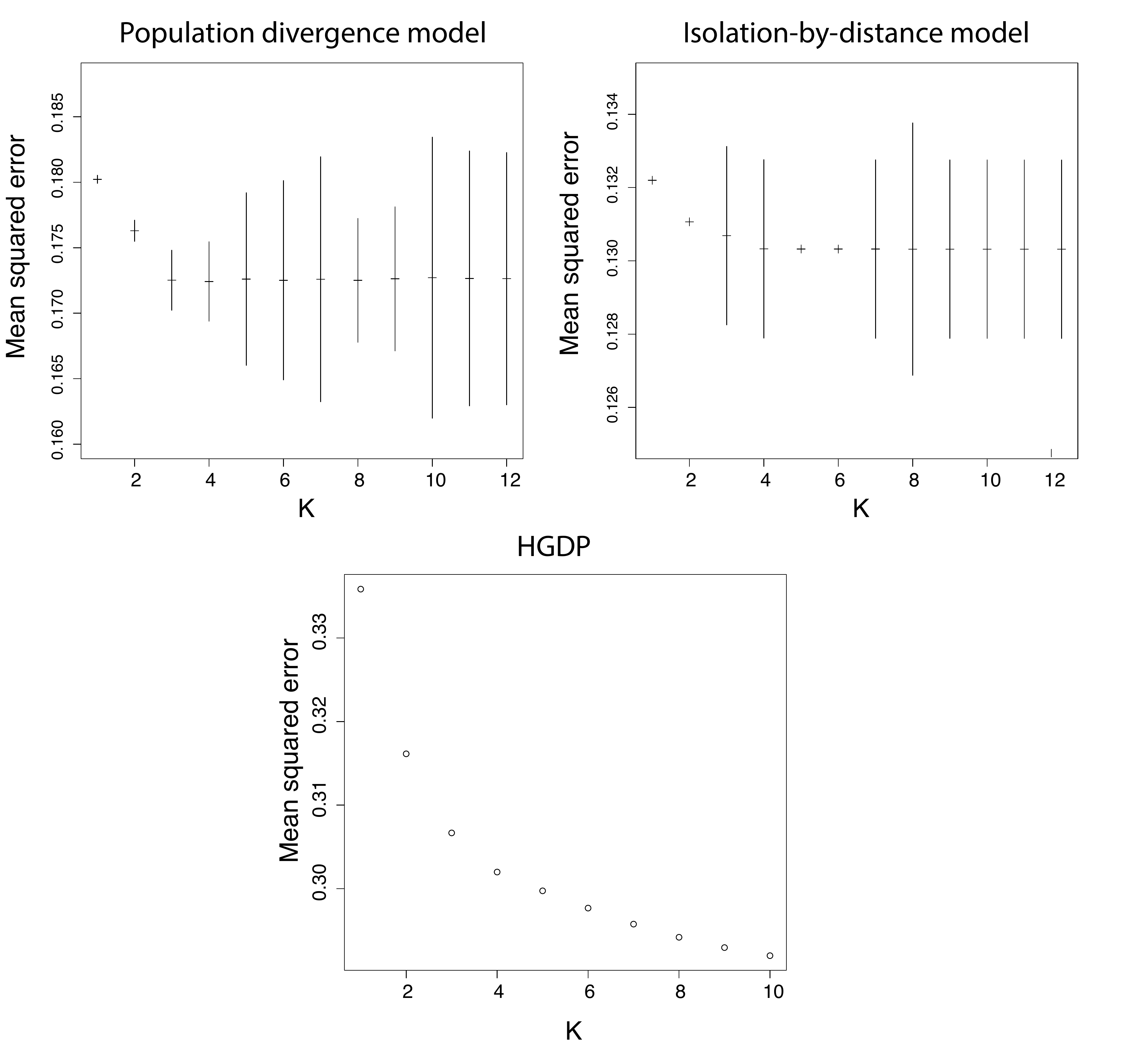}
    \caption{Average mean square error as a function of the number of factors $K$. Error bars correspond to 2 standard deviations and are estimated using a total of 10 different MCMC runs.}
\end{figure}

\clearpage

\begin{figure}
   \includegraphics[scale=.5]{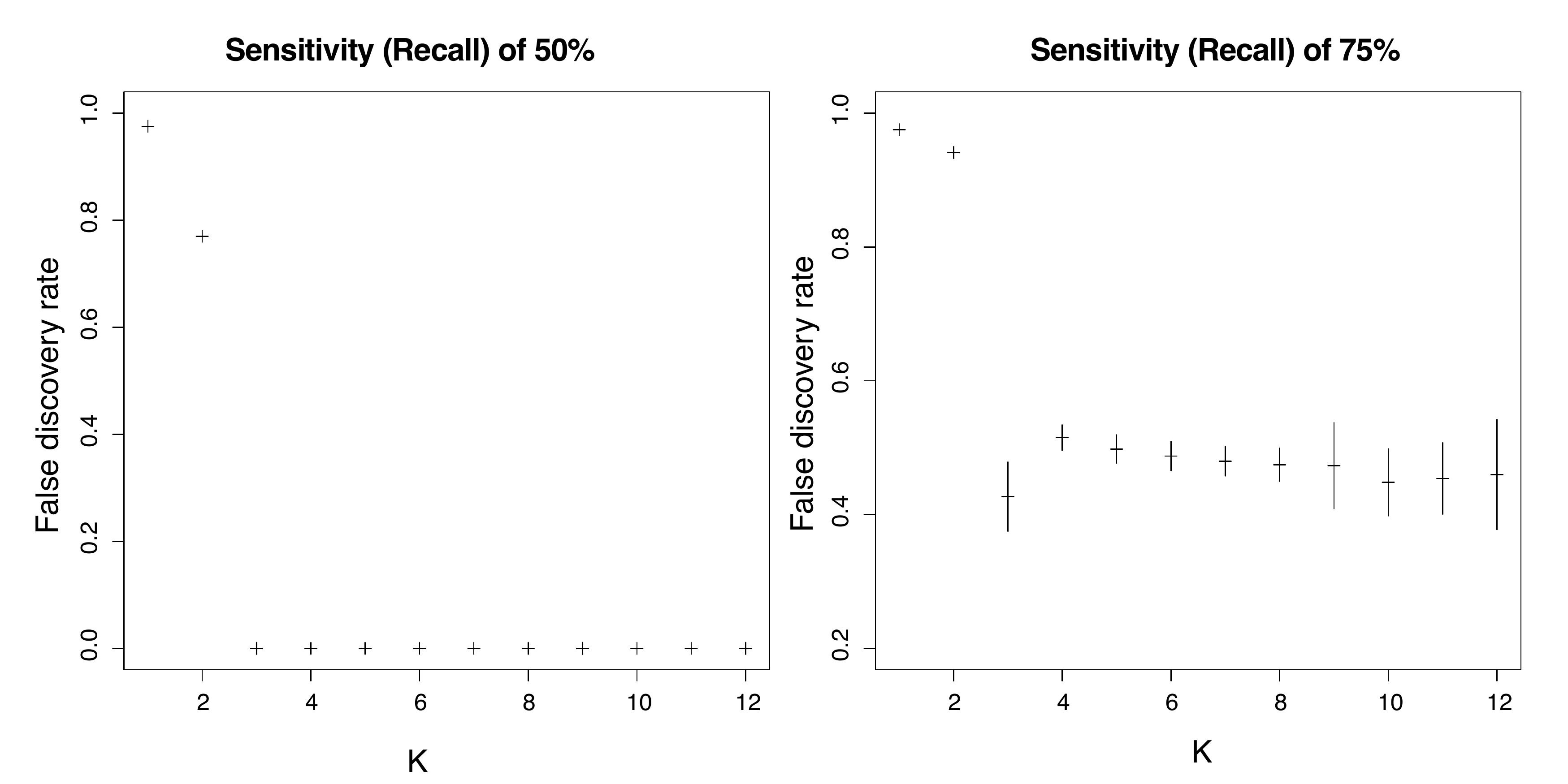}
    \caption{False discovery rate obtained with the factor model using different values of the number of factors $K$ in the IBD model. Error bars correspond to 2 standard deviations and are estimated using a total of 10 different MCMC runs.}
\end{figure}

\clearpage

\begin{figure}
\begin{center}
   \includegraphics[scale=.7]{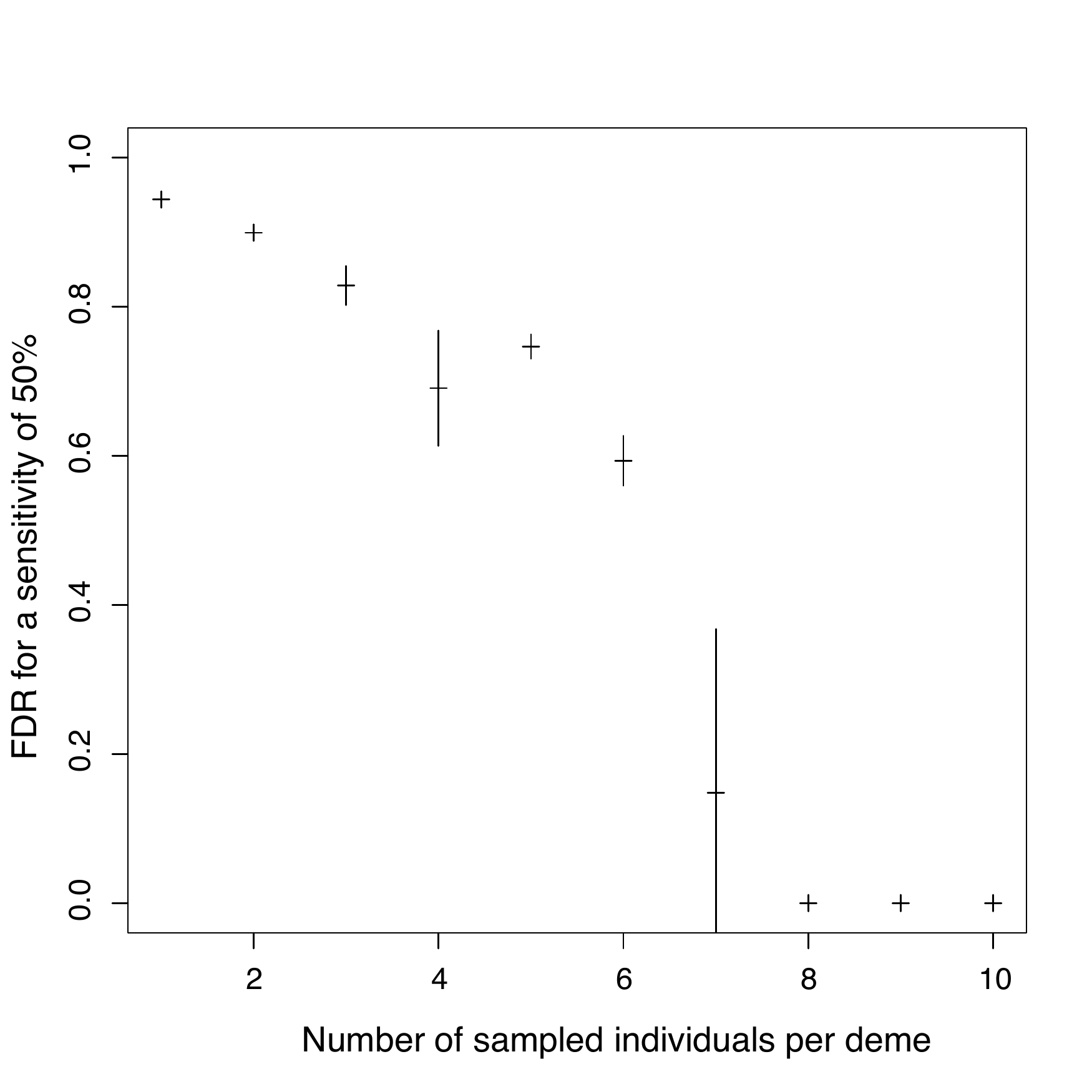}
    \caption{False discovery rate as a function of the number of sampled individuals per deme for the IBD model.}
 \end{center}
\end{figure}

\clearpage

\begin{figure}
   \includegraphics[scale=.5]{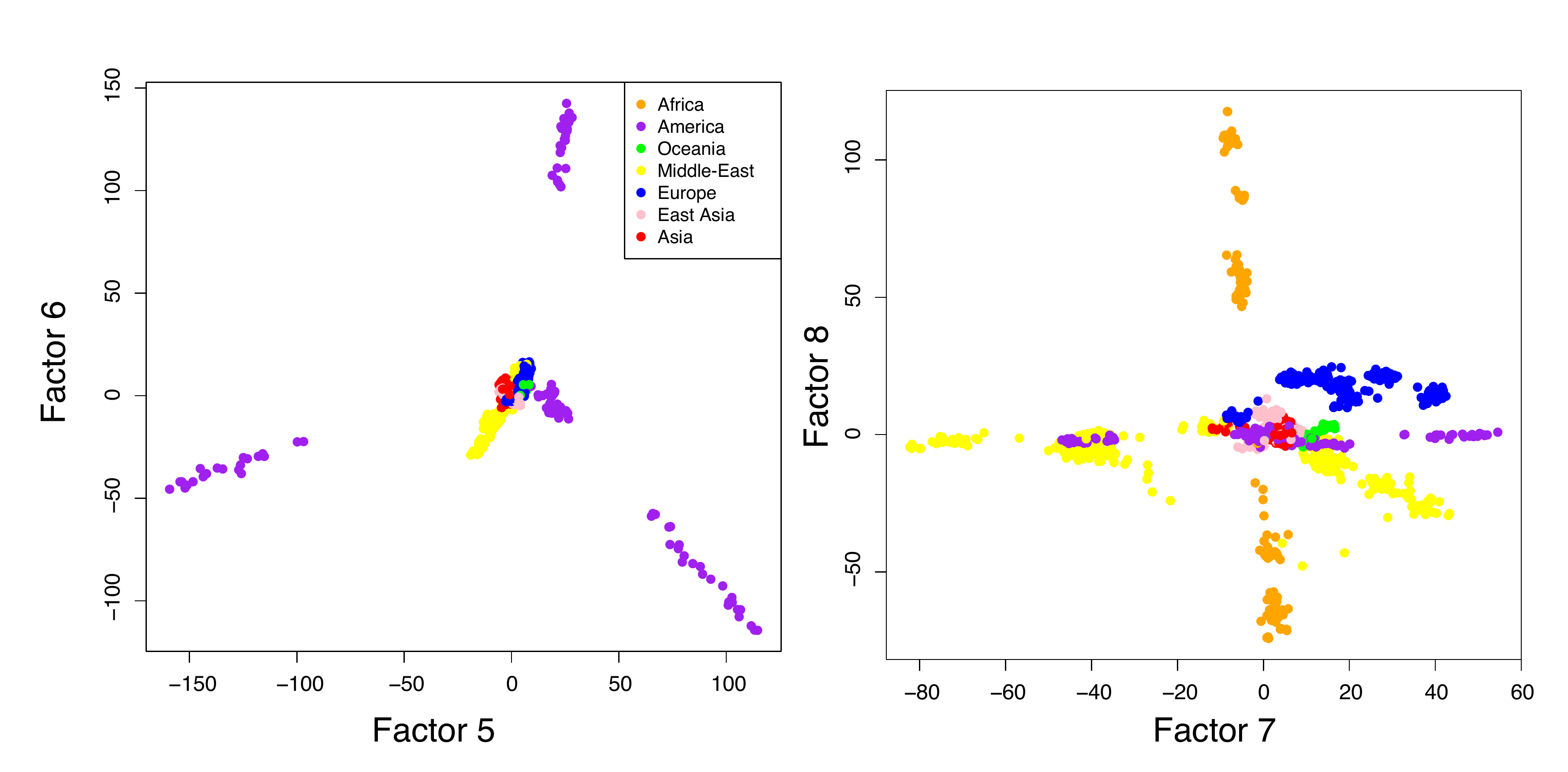}
    \caption{Values of factors $5-8$ for the HGDP dataset.}
\end{figure}

\clearpage

\begin{figure}
   \includegraphics[scale=1]{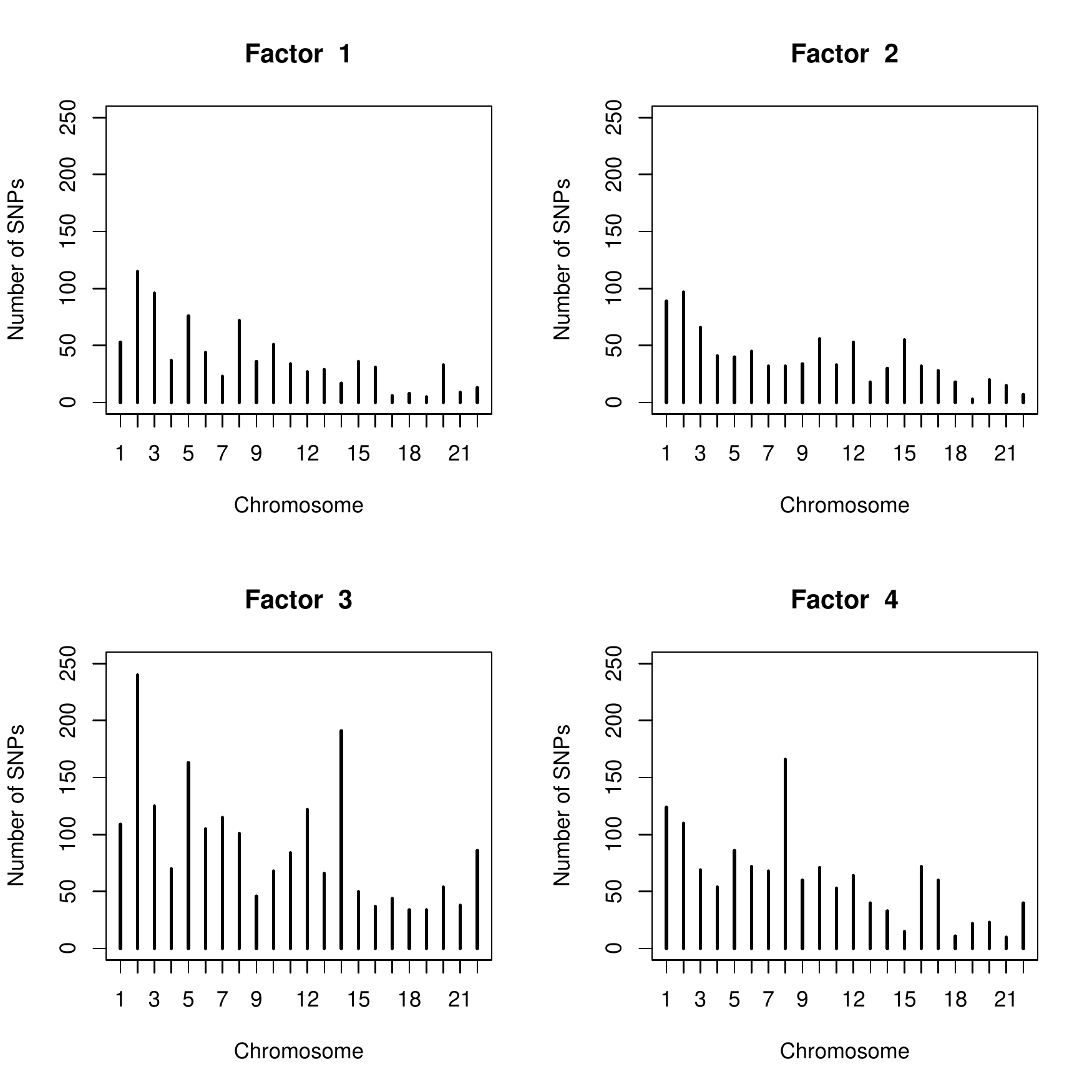}
    \caption{Number of outlier SNPs per chromosome for each of the four different factors.}
\end{figure}

 \addtocounter{table}{1}
\clearpage

\begin{table}
\footnotesize
\centering
\begin{tabular}{ccc}
\hline
rs identifier & Factor & Freq. Map\\ 
\hline
4918664 & 1 & \url{http://hgdp.uchicago.edu/cgi-bin/alfreqs.cgi?pos=94911055&chr=chr10&rs=rs4918664&imp=false}\\ 

10882168 & 1 & \url{http://hgdp.uchicago.edu/cgi-bin/alfreqs.cgi?pos=94919424&chr=chr10&rs=rs10882168&imp=false}\\ 

7091054 & 1 & \url{http://alfred.med.yale.edu/alfred/mvograph.asp?siteuid=SI533583B}\\ 

11187300 & 1 & \url{http://alfred.med.yale.edu/Alfred/mvograph.asp?siteuid=SI091645Z}\\ 

7556886 & 1 & \url{http://alfred.med.yale.edu/alfred/mvograph.asp?siteuid=SI563914C}\\ 

12220128 & 1 & \url{http://alfred.med.yale.edu/alfred/mvograph.asp?siteuid=SI136240Q}\\ 

6583859 & 1 & \url{http://hgdp.uchicago.edu/cgi-bin/alfreqs.cgi?pos=94883463&chr=chr10&rs=rs6583859&imp=false}\\ 

4918924 & 1 & \url{http://alfred.med.yale.edu/alfred/mvograph.asp?siteuid=SI450066V}\\ 

1834619 & 1 & \url{http://alfred.med.yale.edu/alfred/mvograph.asp?siteuid=SI274807C}\\ 

4578856 & 1 & \url{http://alfred.med.yale.edu/alfred/mvograph.asp?siteuid=SI421641S}\\ 

1834640 & 2 & \url{http://hgdp.uchicago.edu/cgi-bin/alfreqs.cgi?pos=46179457&chr=chr15&rs=rs1834640&imp=false}\\ 

2250072 & 2 & \url{http://hgdp.uchicago.edu/cgi-bin/alfreqs.cgi?pos=46172199&chr=chr15&rs=rs2250072&imp=false}\\ 

260714 & 2 & \url{http://alfred.med.yale.edu/alfred/mvograph.asp?siteuid=SI015078V}\\ 

7531501 & 2 & \url{http://alfred.med.yale.edu/alfred/mvograph.asp?siteuid=SI561912Y}\\ 

11637235 & 2 & \url{http://alfred.med.yale.edu/alfred/mvograph.asp?siteuid=SI103044M}\\ 

10760260 & 2 & \url{http://alfred.med.yale.edu/alfred/mvograph.asp?siteuid=SI059055Y}\\ 

2416899 & 2 & \url{http://alfred.med.yale.edu/alfred/mvograph.asp?siteuid=SI333938D}\\ 

2406410 & 2 & \url{http://alfred.med.yale.edu/alfred/mvograph.asp?siteuid=SI333101L}\\ 

3751631 & 2 & \url{http://alfred.med.yale.edu/alfred/mvograph.asp?siteuid=SI386311W}\\ 

618746 & 2 & \url{http://alfred.med.yale.edu/alfred/mvograph.asp?siteuid=SI478518H}\\ 

139553 & 3 & \url{http://alfred.med.yale.edu/alfred/mvograph.asp?siteuid=SI197254C}\\ 

5996039 & 3 & \url{http://alfred.med.yale.edu/alfred/mvograph.asp?siteuid=SI469320Y}\\ 

8139993 & 3 & \url{http://alfred.med.yale.edu/alfred/mvograph.asp?siteuid=SI610594Z}\\ 

126092 & 3 & \url{http://alfred.med.yale.edu/alfred/mvograph.asp?siteuid=SI156639E}\\ 

1005402 & 3 & \url{http://alfred.med.yale.edu/alfred/mvograph.asp?siteuid=SI022498Z}\\ 

8137373 & 3 & \url{http://alfred.med.yale.edu/alfred/mvograph.asp?siteuid=SI610490U}\\ 

133074 & 3 & \url{http://alfred.med.yale.edu/alfred/mvograph.asp?siteuid=SI184162W}\\ 

9611613 & 3 & \url{http://alfred.med.yale.edu/alfred/mvograph.asp?siteuid=SI645506A}\\ 

2424641 & 3 & \url{http://alfred.med.yale.edu/Alfred/mvograph.asp?siteuid=SI334632V}\\ 

2600814 & 3 & \url{http://alfred.med.yale.edu/alfred/mvograph.asp?siteuid=SI345176A}\\ 

16892216 & 4 & \url{http://alfred.med.yale.edu/alfred/mvograph.asp?siteuid=SI231611O}\\ 

6990312 & 4 & \url{http://alfred.med.yale.edu/alfred/mvograph.asp?siteuid=SI525364Z}\\ 

9908046 & 4 & \url{http://alfred.med.yale.edu/alfred/mvograph.asp?siteuid=SI659032Z}\\ 

575873 & 4 & \url{http://alfred.med.yale.edu/alfred/mvograph.asp?siteuid=SI014358V}\\ 

4691075 & 4 & \url{http://alfred.med.yale.edu/alfred/mvograph.asp?siteuid=SI428601V}\\ 

4471745 & 4 & \url{http://alfred.med.yale.edu/alfred/mvograph.asp?siteuid=SI416772B}\\ 

12891534 & 4 & \url{http://alfred.med.yale.edu/alfred/mvograph.asp?siteuid=SI168020R}\\ 

6988341 & 4 & \url{http://alfred.med.yale.edu/alfred/mvograph.asp?siteuid=SI525168B}\\ 

12216712 & 4 & \url{http://alfred.med.yale.edu/alfred/mvograph.asp?siteuid=SI136091U}\\ 

11869714 & 4 & \url{http://alfred.med.yale.edu/alfred/mvograph.asp?siteuid=SI014860T}\\ 
\hline
\end{tabular}
\label{table:table}
\caption{\small{List of the 10 SNPs with largest Bayes factors for each of the four factors obtained with the HGDP dataset. For each SNP, we provide the link to the {\it ALFRED} database to display maps or bar charts of allele frequencies.}} 
\end{table}
\clearpage

\begin{table}
\footnotesize
\centering
\begin{tabular}{llllll}
\hline
Gene Ontology & no. SNP$^{*}$ & no. genes$^{+}$ & on chip$^{-}$ & GO description & Factor\\ 
\hline
GO:0001522 & 6 & 2 & 15 & pseudouridine synthesis & 1\\ 
GO:0030324 & 19 & 5 & 79 & lung development & 1\\ 
GO:0032024 & 9 & 3 & 28 & positive regulation of insulin secretion & 1\\ 
GO:0090277 & 9 & 3 & 35 & positive regulation of peptide hormone secretion & 1\\ 
GO:0046887 & 9 & 3 & 51 & positive regulation of hormone secretion & 1\\ 
GO:0008045 & 11 & 1 & 16 & motor axon guidance & 1\\ 
GO:0051896 & 9 & 4 & 56 & regulation of protein kinase B signaling cascade & 1\\ 
GO:0042993 & 6 & 1 & 18 & positive regulation of transcription factor import into nucleus & 2\\ 
GO:0042346 & 6 & 1 & 14 & positive regulation of NF-kappaB import into nucleus & 2\\ 
GO:0007156 & 61 & 12 & 115 & homophilic cell adhesion & 3\\ 
GO:0009896 & 30 & 4 & 97 & positive regulation of catabolic process & 3\\ 
GO:0031331 & 29 & 3 & 76 & positive regulation of cellular catabolic process & 3\\ 
GO:0000096 & 20 & 1 & 29 & sulfur amino acid metabolic process & 4\\ 
GO:0042982 & 10 & 1 & 13 & amyloid precursor protein metabolic process & 4\\ 
GO:0006516 & 10 & 1 & 15 & glycoprotein catabolic process & 4\\ 
GO:0006555 & 20 & 1 & 17 & methionine metabolic process & 4\\ 
GO:0007568 & 26 & 5 & 175 & aging & 4\\ 
\hline
$^{*}$ number of outlier SNPs\\
$^{+}$ number of outlier genes\\
$^{-}$ number of genes on SNP array

\end{tabular}
\label{table:table}
\caption{\small{Results of the Gene Ontology based on the set of genes found in Table S1. We use a threshold of 0.05 for the false discovery rate, we remove gene ontology terms that are shared by less than 10 genes or more than 1,000 genes. For each factor, we consider the ten GOs with the smallest FDRs and we only report the gene ontology terms that are related to biological processes.}} 
\end{table}

\end{document}